# EXACT TIME-DEPENDENT SOLUTION OF THE SCHRÖDINGER EQUATION, ITS GENERALIZATION TO THE PHASE SPACE AND RELATION TO THE GIBBS DISTRIBUTION

E.E. Perepelkin[a,b,d], B.I. Sadovnikov[a], N.G. Inozemtseva[b,c], I.I. Aleksandrov[a,b]

[a] *Faculty of Physics, Lomonosov Moscow State University, Moscow, 119991 Russia*
[b] *Moscow Technical University of Communications and Informatics, Moscow, 123423 Russia*
[c] *Dubna State University, Moscow region, Dubna,141980 Russia*
[d] *Joint Institute for Nuclear Research, Moscow region, Dubna,141980 Russia*

**Abstract**

Using the simplest but fundamental example, the problem of the infinite potential well, this paper makes an ideological attempt (supported by rigorous mathematical proofs) to approach the issue of «understanding» the mechanism of quantum mechanics processes, despite the well-known examples of the EPR paradox type. The new exact solution of the Schrödinger equation is analyzed from the perspective of quantum mechanics in the phase space. It is the phase space, which has been extensively used recently in quantum computing, quantum informatics and communications, that is the bridge towards classical physics, where understanding of physical reality is still possible. In this paper, an interpretation of time-dependent processes of energy redistribution in a quantum system, probability waves, the temperature and entropy of a quantum system, and the transition to a time-independent «frozen state» is obtained, which is understandable from the point of view of classical physics. The material of the paper clearly illustrates the solution of the problem from the standpoint of continuum mechanics, statistical physics and, of course, quantum mechanics in the phase space.

**Introduction**

The Schrödinger equation is a fundamental equation in quantum mechanics. The presence of an exact solution to the Schrödinger equation greatly simplifies the analysis of a quantum system and has a wide methodological aspect. There is a narrow circle of potentials $U$ for which exact solutions of the Schrödinger equation are known. The simplest problem of quantum mechanics is the problem of finding the probability density for a particle in the infinite potential well:

$$U(x) = \begin{cases} 0, & \text{if } 0 < x < l, \\ +\infty, & \text{otherwise}, \end{cases} \quad \text{(i.1)}$$

where $l$ − is the width of the well.

Despite the seeming simplicity of the problem, this model has a wide range of applications in solid state physics and quantum optics. The particle in a box model (i.1) is used in the search for approximate solutions to more complex physical systems, in which the particle is located in a narrow domain with a low electric potential between two high-potential barriers. Systems with quantum wells [1-3] are important in optoelectronics and are used in devices such as a quantum well laser, an infrared photodetector with quantum wells, and a quantum-confined Stark effect modulator [4,5]. This approach is applicable when considering the lattice in the Kronig-Penney model and for final metal in the approximation of free electrons.

The model of a particle in the infinite potential well can be used when considering quantum dots [6-8]. Quantum dots are very small semiconductors (on the nanometer scale). Thus, a quantum confinement arises in the sense that electrons cannot leave the «dot», which allows one to use the «particle in a box» approximation. The behavior of such systems can be described by three-dimensional energy quantization equations of the «particle in a box» type.

The solution of the Schrödinger equation for potential (i.1):

$$\begin{cases} i\hbar \dfrac{\partial \Psi}{\partial t} = -\dfrac{\hbar^2}{2m}\dfrac{\partial^2 \Psi}{\partial x^2},\ 0 < x < l, \\ \Psi(0,t) = \Psi(l,t) = 0, \end{cases} \quad (i.2)$$

has the form

$$\Psi_\mu(x,t) = \sqrt{\dfrac{2}{l}} \sin\left(\dfrac{\sqrt{2mE_\mu}}{\hbar} x\right) e^{-i\frac{E_\mu}{\hbar}t}, \qquad E_\mu = \dfrac{\pi^2 \hbar^2 \mu^2}{2ml^2},\ \mu \in \mathbb{N}, \quad (i.3)$$

where $\mu$ – is the number of the quantum state. For wave function (i.3) probability density $f_\mu^1$ is a stationary function:

$$f_\mu^1(x) = |\Psi_\mu(x,t)|^2 = \dfrac{2}{l}\sin^2\left(\dfrac{\sqrt{2mE_\mu}}{\hbar} x\right). \quad (i.4)$$

From a mathematical point of view, equation (i.2) as a partial differential equation for its single-valued solution requires specifying not only the boundary conditions (i.2) $\Psi(0,t) = \Psi(l,t) = 0$, but also the initial $\Psi(x,0) = \Psi_0(x)$ or $\Psi_t(x,0) = \Psi_1(x)$. As can be seen from the statement of problem (i.2), the initial condition is not specified. The absence of the initial condition in statement (i.2) is compensated by the choice of the type of solution. Indeed, solution (i.3) was obtained for the particular case when the wave function admits a factorized representation $\Psi(x,t) \sim F(x)T(t)$. Thus, solution (i.3) belongs to a narrow class of possible solutions to problem (i.2).

The purpose of this work is to find new time-dependent solutions of problem (i.2), to construct the Wigner function for them [9, 10] in the phase space and the vector field of the probability flow, to analyze the dynamic properties of such quantum systems from the standpoint of statistical physics (in terms of the Gibbs distribution and entropy), continuum mechanics and quantum mechanics in the phase space.

The paper has the following structure. In §1, an exact $\Psi_{\mu,\beta}(x,t)$ time-dependent solution of problem (i.2) is constructed through the Jacobi $\theta$–function [11, 12], where $\mu$ – is the number of the quantum state, and $\beta$ is interpreted as the thermodynamic parameter of the «inverse temperature» of the quantum system. An explicit expression for probability density $f_{\mu,\beta}^1(x,t)$ is obtained from wave function $\Psi_{\mu,\beta}(x,t)$, which is a periodic function in time with period $T_\mu$ depending on state number $\mu$. An expression is found for function $f_{\mu,\beta}^1(x,t)$ averaged over period $T_\mu$, that is for $\bar{f}_{\mu,\beta}^1(x)$. Theorems are proved that the known stationary distribution density (i.4) is asymptotics at $\beta \to +\infty$ for distributions $f_{\mu,\beta}^1(x,t)$ and $\bar{f}_{\mu,\beta}^1(x)$. A dynamic analysis of the structure of the evolution of distribution function $f_{\mu,\beta}^1(x,t)$ at various values of inverse temperature $\beta$ is carried out.

In §2, a quantum system is considered in the phase space. An explicit expression is constructed for the Wigner function $f_{\mu,\beta}^{1,2}(x,v,t) = \hbar W_{\mu,\beta}(x,p,t)$, which satisfies the evolutionary Moyal equation [13] (a quantum analogue of the Liouville equation). The connection between the Moyal equation and the second Vlasov equation [14] through the Vlasov-Moyal approximation [15] for the acceleration field of probability flux $\langle \dot{\vec{v}} \rangle$ is described. Using the first Vlasov equation for function $f_{\mu,\beta}^1(x,t)$, the velocity field of the probability flow $\langle \vec{v} \rangle$ is found in an explicit form. Periodic probability waves that arise inside the potential well at low values of inverse temperature $\beta$ and their damping with decreasing temperature («freezing» of the quantum system) are considered.

In §3, a thermodynamic model of a quantum system described by the Gibbs distribution is constructed. Period-averaged probability density function $\bar{f}_{\mu,\beta}^1(x)$ is represented as the Gibbs average of the ensemble of wave packets of all frequencies. The Gibbs averaged value of energies $\langle \mathcal{E}_\mu \rangle_{Gibbs}(\beta)$ of a quantum system has asymptotics at $\beta \to +\infty$ in the form of eigenvalue spectrum $E_\mu$ of the time-independent system (i.3). The dynamic process of energy redistribution inside the potential well depending on the «temperature» of the quantum system is considered in detail. An explicit expression for the thermodynamic entropy of a quantum system is obtained. Entropy is a strictly monotonic function of the «temperature» of a quantum system. A theorem is proved about the tendency of thermodynamic entropy to zero during freezing of a quantum system. From the perspective of continuum mechanics, it is shown that the laws of conservation of «mass» (probability), momentum (probability flux density) and energy are satisfied.

The Conclusions contain a generalization of the main results of the paper. The Appendix contains proofs of the theorems.

### §1 Exact solution of the Schrödinger equation

Let us construct a time-dependent exact solution of the Schrödinger equation to the boundary value problem (i.2), which is different from solution (i.3).

**Theorem 1.** *Wave function $\Psi_{\mu,\beta}(x,t)$ is a time-dependent solution of the boundary value problem (i.2)*

$$\Psi_{\mu,\beta}(x,t) = \frac{1}{\sqrt{N(\beta)}} \theta_1\left(\mu \frac{x}{l}, -\mu^2 \frac{2\pi\hbar}{ml^2} t + i\beta\right), \qquad (1.1)$$

$$N(\beta) = l \sum_{k=-\infty}^{+\infty} e^{-\frac{\pi\beta}{2}(2k+1)^2}, \qquad (1.2)$$

$$\Psi_{\mu,\beta}(0,t) = \Psi_{\mu,\beta}(l,t) = 0, \quad \int_0^l |\Psi_{\mu,\beta}(x,t)|^2 \, dx = 1.$$

*where $\mu-$ is the number of the quantum state; parameter $\beta > 0$; $\theta_1-$ is the Jacobi theta-function*

$$\theta_1(z,\tau) = \sum_{k=-\infty}^{+\infty} e^{i\frac{\pi\tau}{4}(2k+1)^2 + i\frac{\pi}{2}(2z+1)(2k+1)}, \quad z \in \mathbb{C}, \ \tau = \alpha + i\beta, \ \alpha, \beta \in \mathbb{R}. \qquad (1.3)$$

The proof of Theorem 1 is given in the Appendix.

**Remark** Solution $\Psi_{\mu,\beta}(x,t)$ can be represented in the form

$$\Psi_{\mu,\beta}(x,t) = \frac{1}{\sqrt{N(\beta)}} \theta_1\left(\frac{\sqrt{2m\varepsilon_\mu}}{\hbar} x, -\frac{4\pi\varepsilon_\mu}{\hbar} t + i\beta\right), \tag{1.4}$$

where

$$\varepsilon_\mu = \frac{\hbar^2 \mu^2}{2ml^2}, \quad E_\mu = \pi^2 \varepsilon_\mu. \tag{1.5}$$

Wave function $\Psi_{\mu,\beta}(x,t)$ differs significantly from the known wave function (i.3). In contrast to function $f_\mu^1(x)$, probability density function $f_{\mu,\beta}^1(x,t)$ for wave function (1.1) will be time-dependent.

**Theorem 2.** *Probability density function $f_{\mu,\beta}^1(x,t)$ for wave function $\Psi_{\mu,\beta}(x,t)$ is periodical and has the form*

$$f_{\mu,\beta}^1(x,t) = |\Psi_{\mu,\beta}(x,t)|^2 = \frac{1}{N(\beta)} \sum_{n,k=-\infty}^{+\infty} e^{-\frac{\pi\beta}{4}\left[(2k+1)^2 + (2n+1)^2\right]} T_{|k-n|}\left(\cos\left[\vartheta_{n,k}^\mu(x,t)\right]\right), \tag{1.6}$$

$$\vartheta_{n,k}^\mu(x,t) \stackrel{det}{=} \pi\left(2\mu\frac{x}{l} + 1\right) - \frac{\pi t}{T_\mu}(n+k+1), \tag{1.7}$$

$$T_\mu = \frac{\pi\hbar}{4E_\mu} = \frac{ml^2}{2\pi\hbar\mu^2}, \tag{1.8}$$

*where $T_\mu$ – is the period of $f_{\mu,\beta}^1(x,0) = f_{\mu,\beta}^1(x,T_\mu)$; $T_{|k-n|}$ – are the Chebyshev polynomials.*

The proof of Theorem 2 is given in the Appendix.

**Remark** Function $\vartheta_{n,k}^\mu(x,t)$ can be considered as the characteristic of $\vartheta_{n,k}^\mu(x,t) = const$, along which the corresponding term of functional series (1.6) is constant. The characteristics of (1.7) are straight lines

$$\frac{x}{l} = \frac{n+k+1}{2\mu} \frac{t}{T_\mu} + const \Rightarrow \operatorname{tg} \gamma = \frac{n+k+1}{2\mu}, \tag{1.9}$$

where $\gamma$ – is the characteristic angle. According to expression (1.9), the terms of the series for which $n+k = const$ have characteristics with the same angle $\gamma$.

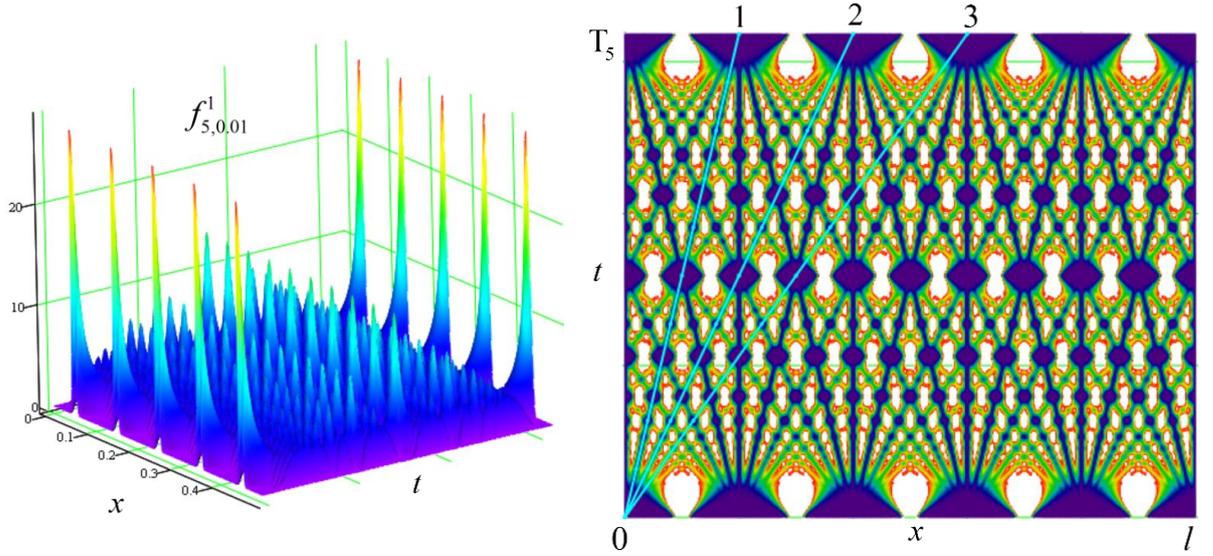

Fig. 1 Evolution of probability density distribution $f^1_{5,0.01}$.

Of course, the characteristics of the various terms of the series overlap, but the main contribution is made by the terms of series (1.6), which have the greatest weighting coefficient $e^{-\frac{\pi\beta}{4}\left[(2k+1)^2+(2n+1)^2\right]}$.

Fig. 1 shows the evolution of distribution density $f^1_{\mu,\beta}(x,t)$ for quantum state $\mu=5$ and parameter $\beta=0.01$. Time $t$ in Fig. 1 varies within one period $0\leq t\leq T_5$ (1.8) and coordinate $x: 0\leq x\leq l$. The left-hand side of Fig. 1 shows an isometric distribution projection $f^1_{5,0.01}(x,t)$, and the right-hand side shows the level lines of function $f^1_{5,0.01}(x,t)$ when viewed from above onto plane $(x,t)$. In Fig. 1 (right), characteristic straight lines (1.9) numbered «1», «2» and «3» are shown in blue with parameters $n+k=0$, $n+k=1$ and $n+k=2$, respectively.

It is seen that characteristics (1.9) define the basic structure of the evolution of the probability distribution density at small $\beta$. With increasing $\beta$, the main contribution to series (1.6) is made by the summands corresponding to the values of indices $n+k+1=0$. In this case, the period (1.8) tends to infinity, and characteristics (1.9) turn into vertical lines along the time axis. Let us formulate the following theorem.

**Theorem 3.** *Time-independent probability density function $f^1_\mu(x)$ (i.4) is the asymptotics for the time-dependent function $f^1_{\mu,\beta}(x,t)$ (1.6) for large values of the parameter $\beta$, that is*

$$\lim_{\beta\to+\infty} f^1_{\mu,\beta}(x,t) = f^1_\mu(x). \qquad (1.10)$$

The proof of Theorem 3 is given in the Appendix.

Fig. 2 shows the evolutions of distribution functions $f^1_{5,\beta}(x,t)$ for large values of $\beta$. In Fig. 2, it can be seen that with an increase in parameter $\beta$, the inhomogeneous «time structure» of the probability distribution begins to blur and function $f^1_{5,\beta}(x,t)$ tends to time-independent distribution $f^1_{5,\beta}(x,t)\to f^1_5(x)$ (i.4).

Let us perform averaging over time (over period $T_\mu$) of distribution function (1.6).

$$\overline{f}^1_{\mu,\beta}(x) \stackrel{det}{=} \frac{1}{T_\mu} \int_0^{T_\mu} f^1_{\mu,\beta}(x,t)\,dt. \qquad (1.11)$$

**Theorem 4.** *For function $\overline{f}^1_{\mu,\beta}$, the following representation is valid*

$$\overline{f}^1_{\mu,\beta}(x) = \frac{2}{N(\beta)} \sum_{k=-\infty}^{+\infty} e^{-\frac{\pi\beta}{2}(2k+1)^2} \sin^2\left[(2k+1)\frac{\pi\mu}{l}x\right], \qquad (1.12)$$

*where* $\dfrac{\pi\mu}{l} = \dfrac{\sqrt{2mE_\mu}}{\hbar}$.

The proof of Theorem 4 is given in the Appendix.

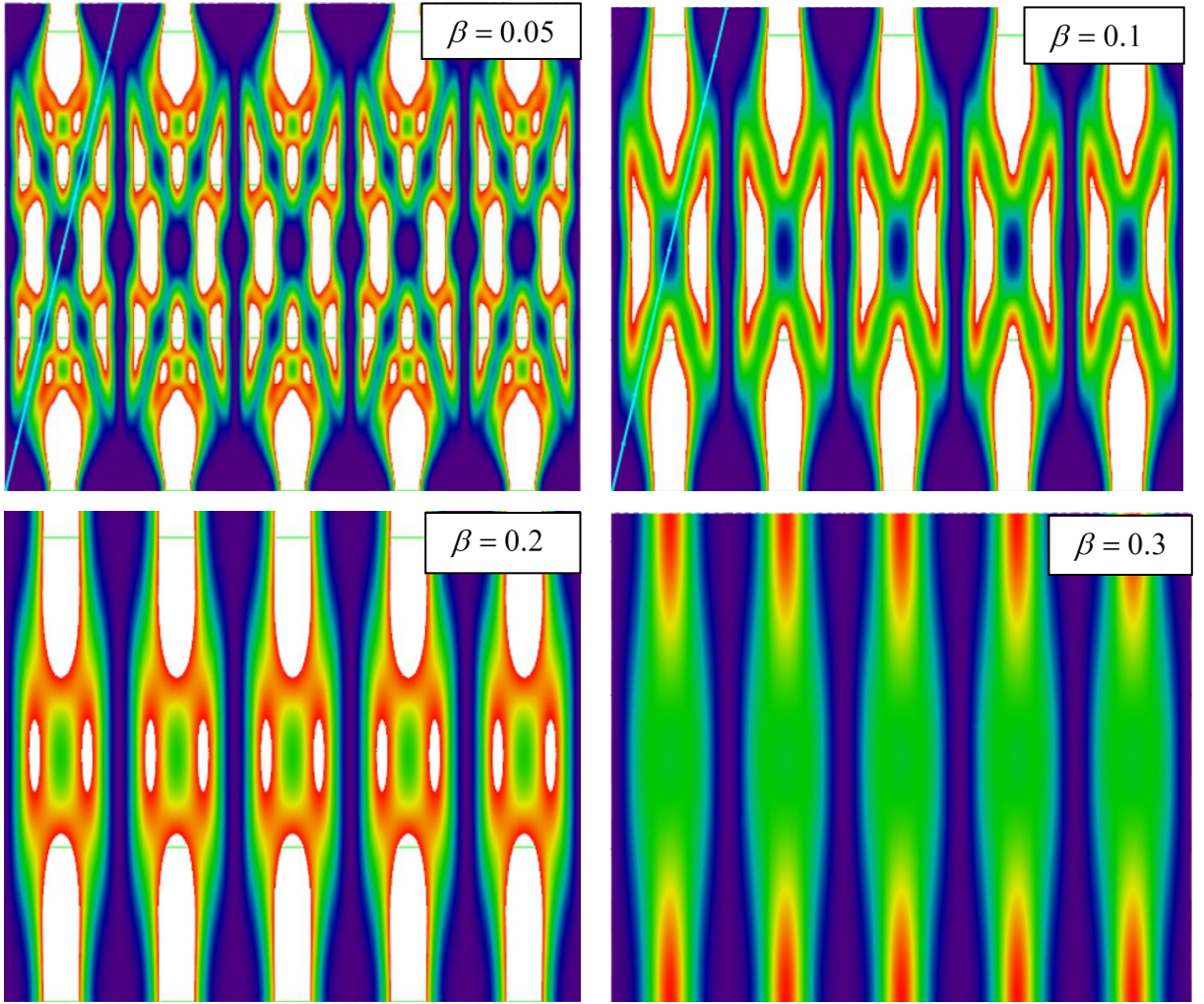

Fig. 2 Blurring of the «time inhomogeneity» of $f^1_{5,\beta}$ with an increase of $\beta$.

Let us see how distributions (i.4) and (1.12) are related. Fig. 3 shows the comparison of distributions $\overline{f}^1_{5,\beta}(x)$ and $f^1_{5,\beta}(x,0)$ to distribution $f^1_5$ for various values of parameter $\beta$. On the top of Fig. 3 the comparison of distribution $f^1_{5,\beta}(x,0)$ to distribution $f^1_5$ is shown, and the bottom part of the Fig.3 shows the comparison of $\overline{f}^1_{5,\beta}$ to $f^1_5$.

By analogy to Figs. 1, 2, Fig. 3 illustrates that for small values of $\beta$ new distribution density $f^1_{5,\beta}(x,0)$ differs significantly from the known distribution density $f^1_5$, but with an increase of $\beta$ the difference gets smaller. Averaged function $\overline{f}^1_{5,\beta}(x)$ at $\beta \to 0$ degenerates to a Dirac $\delta$-function, which leads to the so called «Dirac comb», but with increasing $\beta$ averaged function $\overline{f}^1_{5,\beta}$ converges to $f^1_5$.

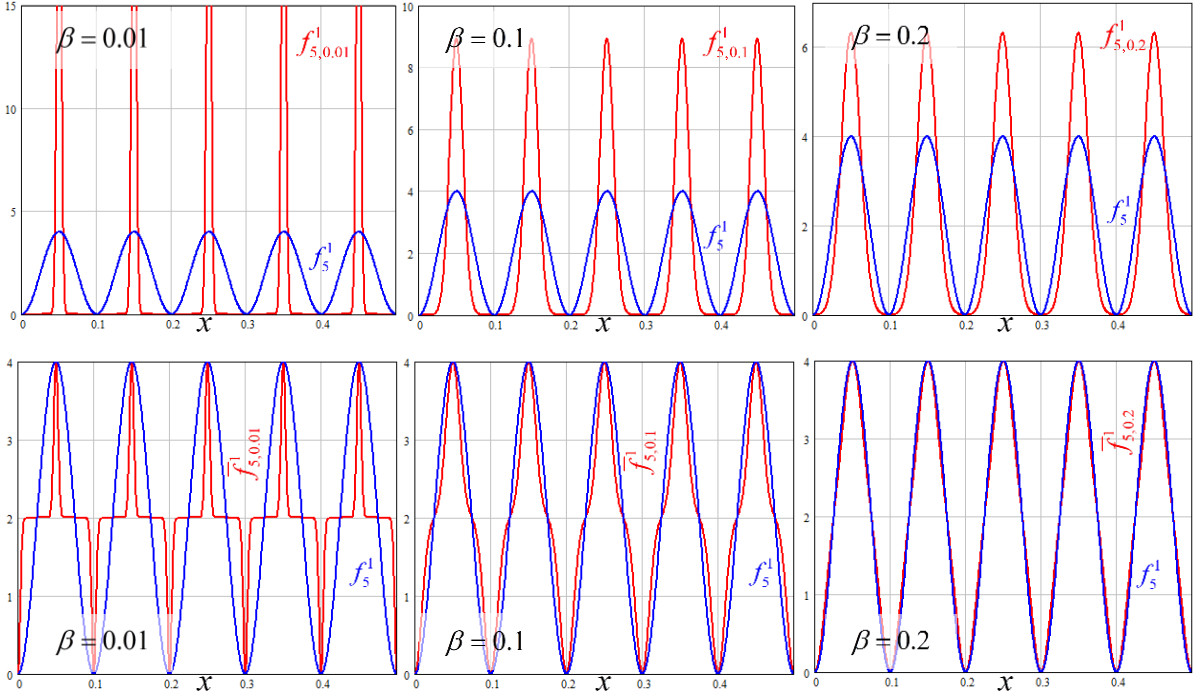

Fig. 3 Comparison of distributions $f^1_{5,\beta}(x,0)$ and $\overline{f}^1_{5,\beta}(x)$ with distribution $f^1_5(x)$.

This numerical simulation result may be formulated as a theorem.

**Theorem 5.** *Time-independent probability density function $f^1_\mu(x)$ (i.4) is the asymptotics for the time-averaged distribution $\overline{f}^1_{\mu,\beta}(x)$ (1.12) for large values of parameter $\beta$, that is*

$$\lim_{\beta \to +\infty} \overline{f}^1_{\mu,\beta}(x) = f^1_\mu(x). \qquad (1.13)$$

The proof of Theorem 5 follows directly from the integration (1.11) of expression (1.6) over period $T_\mu$.

**§2 The Wigner function**

Let us find the Wigner function $f^{1,2}_{\mu,\beta}(x,v,t) = \hbar W_{\mu,\beta}(x,p,t)$ [9, 10] corresponding to wave function (1.1)

$$W_{\mu,\beta}(x,p,t) = \frac{1}{2\pi\hbar} \int_{-\infty}^{+\infty} \Psi_{\mu,\beta}^{*}\left(x-\frac{s}{2},t\right)\Psi_{\mu,\beta}\left(x+\frac{s}{2},t\right) e^{-i\frac{ps}{\hbar}} ds. \qquad (2.1)$$

**Theorem 6.** *The Wigner function $W_{\mu,\beta}(x,p,t)$ (2.1) for the time-dependent solution of the Schrödinger equation (1.1) has the form:*

$$W_{\mu,\beta}(x,p,t) = \frac{1}{\hbar N(\beta)} \sum_{n,k=-\infty}^{+\infty} e^{-\frac{\pi\beta}{4}\left[(2n+1)^{2}+(2k+1)^{2}\right]} \delta\left(P_{n,k}^{\mu} - p\right) T_{|k-n|}\left[\cos\vartheta_{n,k}^{\mu}(x,t)\right], \qquad (2.2)$$

*where* $\vartheta_{n,k}^{\mu}(x,t) = \frac{2\pi\mu}{l}\left(-\frac{P_{n,k}^{\mu}}{m}t + x\right) + \pi$; $P_{n,k}^{\mu} \stackrel{\text{det}}{=} \sqrt{2mE_{\mu}}(n+k+1)$; $\delta\left(P_{n,k}^{\mu} - p\right) -$ *is the Dirac delta-function;* $T_{|k-n|} -$ *are the Chebyshev polynomials.*

The proof of the theorem is given in the Appendix.

**Remark** From expression (2.2) it is seen that the Wigner function is represented in the form of a superposition of point sources of probability density. Integrating the Wigner function (2.2) over momentum space gives probability density $f_{\mu,\beta}^{1}(x,t)$, that is

$$f_{\mu,\beta}^{1}(x,t) = \int_{-\infty}^{+\infty} f_{\mu,\beta}^{1,2}(x,v,t) dv = \hbar \int_{-\infty}^{+\infty} W_{\mu,\beta}(x,p,t) dp =$$
$$= \frac{1}{N(\beta)} \sum_{n,k=-\infty}^{+\infty} e^{-\frac{\pi\beta}{4}\left[(2n+1)^{2}+(2k+1)^{2}\right]} T_{|k-n|}\left[\cos\vartheta_{n,k}^{\mu}(x,t)\right], \qquad (2.3)$$

which coincides with expression (1.6).

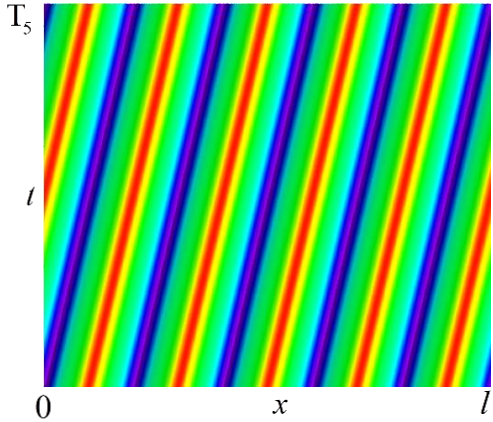

Fig. 4 Evolution of «point» source of quasi-density of probabilities

Each «point» source in expression (2.2) with momentum $p = P_{n,k}^{\mu}$ will consist of a set of harmonics moving along one characteristic $\vartheta_{n,k}^{\mu}(x,t) = const$. The angle of inclination $\gamma$ of such characteristics (1.9) will be determined by the momentum $p = P_{n,k}^{\mu}$. According to Hudson's theorem [16], function (2.2) will have domains of negative values, which justifies its name as a function of quasi-probabilities [17, 18]. The evolution of the quasi-density of probabilities of such a «point» source (at $n+k=1$) will have the form shown in Fig. 4. The red color in Fig. 4 corresponds to the maximum value, and the blue corresponds to the minimum value of the function. As seen from Fig. 4, the area of the phase domain in which the Wigner function is negative (blue) remains constant over time, which corresponds to Theorem 7 from [19].

The Wigner function (2.2) admits the limit transition at $\beta \to +\infty$ and tends to the time-independent Wigner function for wave function (i.3).

Let us calculate the vector field of a probability flux. The first Vlasov equation for the system under consideration will have the form [19, 20]:

$$\frac{\partial f^1_{\mu,\beta}}{\partial t} + \langle v \rangle^{\mu,\beta}_1 \frac{\partial f^1_{\mu,\beta}}{\partial x} + f^1_{\mu,\beta} \frac{\partial \langle v \rangle^{\mu,\beta}_1}{\partial x} = 0, \qquad (2.4)$$

or

$$\frac{\partial \langle v \rangle^{\mu,\beta}_1}{\partial x} + \frac{\partial S^1_{\mu,\beta}}{\partial x} \langle v \rangle^{\mu,\beta}_1 + \frac{\partial S^1_{\mu,\beta}}{\partial t} = 0, \quad S^1_{\mu,\beta} \stackrel{det}{=} \ln f^1_{\mu,\beta}, \qquad (2.5)$$

where $\langle v \rangle^{\mu,\beta}_1$ − is the average velocity of probability flux determined from the second Vlasov equation [19] for functions $f^{1,2}_{\mu,\beta}(x,v,t)$

$$\frac{\partial f^{1,2}_{\mu,\beta}}{\partial t} + v \frac{\partial f^{1,2}_{\mu,\beta}}{\partial x} + \frac{\partial}{\partial v}\left[ f^{1,2}_{\mu,\beta} \langle \dot{v} \rangle^{\mu,\beta}_{1,2} \right] = 0, \qquad (2.6)$$

$$\langle v \rangle^{\mu,\beta}_1(x,t) \stackrel{det}{=} \frac{\int_{-\infty}^{+\infty} f^{1,2}_{\mu,\beta}(x,v,t) v \, dv}{\int_{-\infty}^{+\infty} f^{1,2}_{\mu,\beta}(x,v,t) \, dv} = \frac{\int_{-\infty}^{+\infty} f^{1,2}_{\mu,\beta}(x,v,t) v \, dv}{f^1_{\mu,\beta}(x)}, \qquad (2.7)$$

Equation (2.6) transforms into the Moyal equation [13] for the Wigner function $W(x,p,t)$ with the Vlasov-Moyal approximation for the average acceleration flux $\langle \dot{v} \rangle^{\mu,\beta}_{1,2}$ [15]:

$$\langle \dot{v} \rangle^{\mu,\beta}_{1,2}(x,v,t) = \sum_{k=0}^{+\infty} \frac{(-1)^{k+1}(\hbar/2)^{2k}}{m^{2k+1}(2k+1)!} \frac{\partial^{2k+1} U}{\partial x^{2k+1}} \frac{1}{f^{1,2}_{\mu,\beta}} \frac{\partial^{2k} f^{1,2}_{\mu,\beta}}{\partial v^{2k}} = 0, \qquad (2.8)$$

where the type of potential $U$ (i.1) is taken into consideration. Substituting (2.8) into equation (2.6), we obtain

$$\frac{\partial f^{1,2}_{\mu,\beta}}{\partial t} + v \frac{\partial f^{1,2}_{\mu,\beta}}{\partial x} = 0. \qquad (2.9)$$

On the one hand, equation (2.9) corresponds to the Moyal equation for the Wigner function of a quantum system with potential (i.1), and on the other hand, equation (2.9) is a transport equation, the solution of which may be found by the method of characteristics [26]:

$$f^{1,2}_{\mu,\beta}(x,t) = F_{\mu,\beta}(\xi(x,t)), \qquad \xi(x,t) = x - vt, \qquad (2.10)$$

where $\xi(x,t) = const$ − is a characteristic, and $F_{\mu,\beta}(\xi)$ − is some function determined from initial-boundary conditions or as the Wigner function (2.2) of wave function $\Psi_{\mu,\beta}(x,t)$ (1.4).

Function $\vartheta^{\mu}_{n,k}(x,t)$ in expression (2.2) actually determines characteristic $\xi(x,t)$ in transport equation (2.9), (2.10), since the $\delta$−function will contribute to expression (2.2) only at $p = mv = P^{\mu}_{n,k}$.

As a result, velocity distribution $\langle \vec{v} \rangle_1^{\mu,\beta}(x,t)$ may be obtained in two ways. The first way is to perform the integration according to the formula (2.7) using the expression for the Wigner function (2.2). The second way is to solve the first Vlasov equation (2.4) / (2.5) with respect to $\langle \vec{v} \rangle_1^{\mu,\beta}(x,t)$ knowing distribution function $f_{\mu,\beta}^1(x,t)$. In both cases, the result will be the same. Indeed, the first way gives the expression

$$\langle v \rangle_1^{\mu,\beta}(x,t) = \frac{l}{2\mu T_\mu N(\beta) f_{\mu,\beta}^1(x)} \sum_{n,k=-\infty}^{+\infty} e^{-\frac{\pi\beta}{4}\left[(2n+1)^2+(2k+1)^2\right]} (n+k+1) T_{|k-n|}\left[\cos \vartheta_{n,k}^\mu(x,t)\right]. \quad (2.11)$$

The solution in the second way follows from the following theorem.

**Theorem 7** *The solution to the first Vlasov equation (2.4)/(2.5) can be represented as*

$$\langle v \rangle_1^{\mu,\beta}(x,t) = \frac{l}{2\mu T_\mu f_{\mu,\beta}^1(x,t) N(\beta)} \times$$
$$\times \sum_{n,k=-\infty}^{+\infty} e^{-\frac{\pi\beta}{4}\left[(2k+1)^2+(2n+1)^2\right]} (n+k+1) \cos\left[\vartheta_{n,k}^\mu(x,t)(k-n)\right] + \frac{const}{f_{\mu,\beta}^1(x,t)}, \quad (2.12)$$

*where value*

$$\langle\langle v \rangle\rangle(t) = \int_0^l f_{\mu,\beta}^1(x,t) \langle v \rangle_1^{\mu,\beta}(x,t) dx = const \cdot l, \quad (2.13)$$

*corresponds to the velocity of the center of mass motion and the following asymptotics is valid*

$$\lim_{\beta \to +\infty} \langle v \rangle_1^{\mu,\beta}(x,t) = 0. \quad (2.14)$$

The proof of Theorem 7 is given in the Appendix.

**Remark** From expression (2.13) it follows that the value $const \cdot l$ determines the constant velocity of the center of mass motion of a quantum system with potential (i.1). Without limiting the generality, for further consideration we will assume value (2.13) to be equal to zero. As a result, the solution obtained by the second method (2.12) completely coincides with the solution (2.11) obtained by the first method.

Let us consider a state with number $\mu = 1$ and parameter $\beta = 0.1$, which illustrates the evolution of velocity field distribution $\langle v \rangle_1^{1,0.1}$ (2.11). By analogy with Fig. 1, Fig. 5 shows the evolution of probability density function $f_{1,0.1}^1$. On the left in Fig. 5 an isometric projection is shown and the right-hand side of the Fig. 5 shows a top view with two main characteristics (1.9).

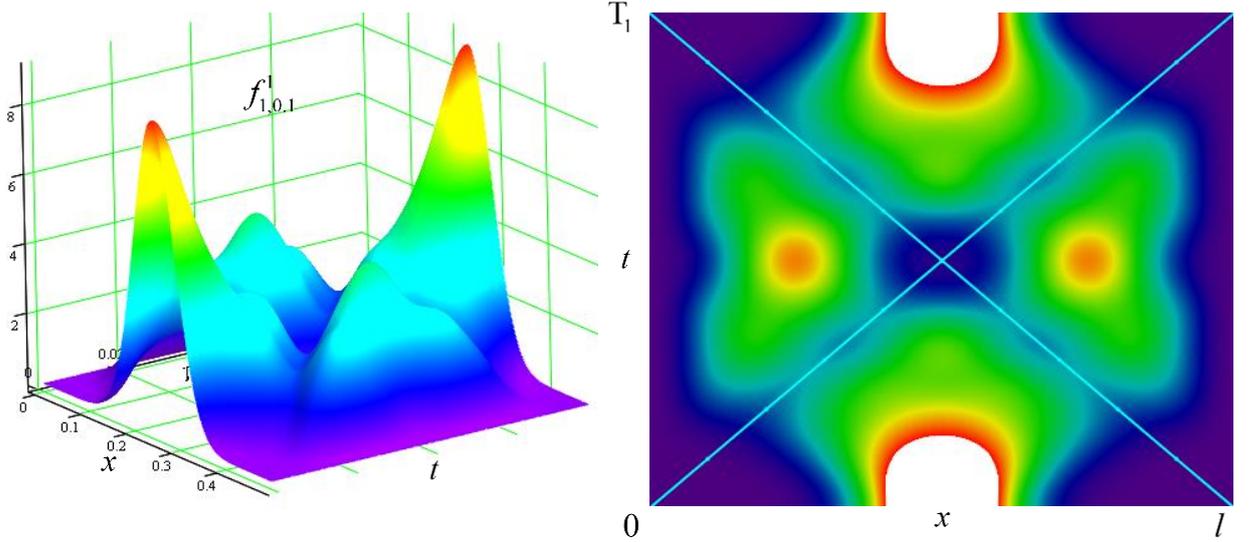

Fig. 5 Evolution of the distribution of probability density $f_{1,0.1}^1$.

Fig. 6 shows a comparison of probability density distributions: $f_1^1(x)$ − time-independent solution (i.4); $f_{1,0.1}^1(x,0)$ − new time-dependent solution (1.6) at the initial moment of time; $\overline{f}_{1,0.1}^1(x)$ − new time-dependent solution (1.12) averaged over the period.

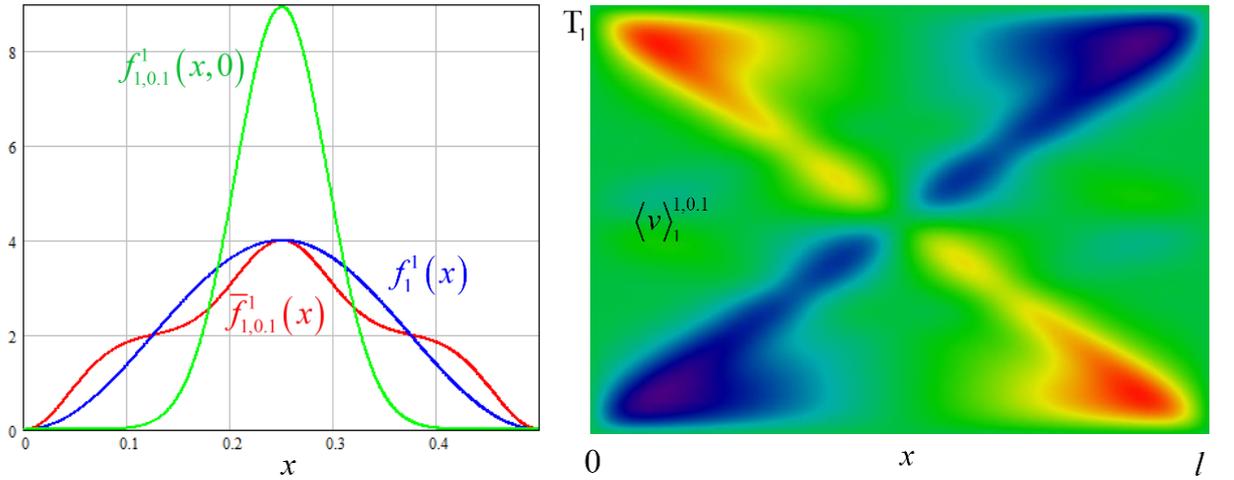

Fig. 6 Distributions of probability densities for $\mu = 1$ and $\beta = 0.1$

Fig. 7 Evolution of the distribution of velocity field $\langle v \rangle_1^{1,0.1}$

In Fig. 6, it can be seen that due to the smallness of parameter $\beta$, the graphs differ significantly from each other. A similar situation was observed earlier for the state with number $\mu = 5$ (see Fig. 3). Fig. 7 shows a top view of the evolution of the distribution of velocity field $\langle v \rangle_1^{1,0.1}$. The blue color corresponds to the minimum values and the maximum values are red. As seen in Fig. 7, the evolution of velocity field $\langle v \rangle_1^{1,0.1}$ has an oscillatory character. At the edges of the potential well ($x = 0$ and $x = l$), velocity $\langle v \rangle_1^{1,0.1}$ is zero (green in Fig. 7) at all times. Inside the potential well, a rapid propagation of probability waves initially occurs to the right (red) and to the left (blue) relative to the center of the potential well (see Fig. 7).

During the first half-period $t \in (0, T_1/2)$, the probability waves, being reflected from the walls of the potential well, gradually fade away and reflected waves appear, propagating in opposite directions during the second half-period $t \in (T_1/2, T_1)$ (see Fig. 7). At the end of period $t = T_1$, as well as at its beginning, the velocities are equal to zero.

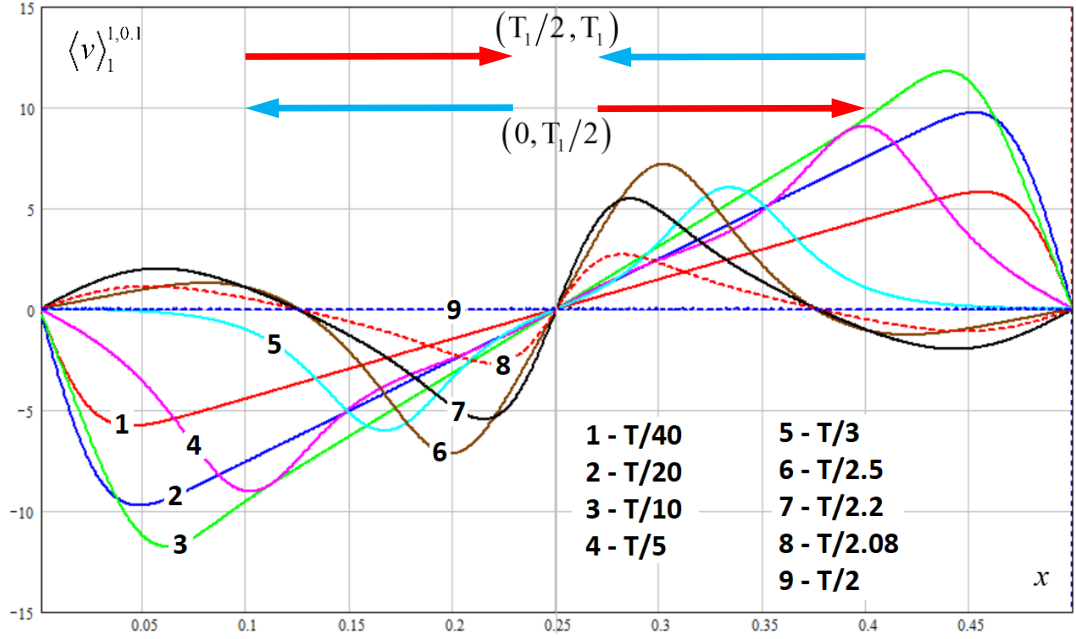

Fig. 8 Evolution of the distribution of velocity field $\langle v \rangle_1^{1,0.1}$ during half-period $t \in (0, T_1/2)$

Fig. 8 shows «time slices» of distribution $\langle v \rangle_1^{1,0.1}$ (see Fig. 7) in the instants: $t_1 = T_1/40$, $t_2 = T_1/20$, $t_3 = T_1/10$, $t_4 = T_1/5$, $t_5 = T_1/3$, $t_6 = T_1/2.5$, $t_7 = T_1/2.2$, $t_8 = T_1/2.08$, $t_9 = T_1/2$.

The evolution of the propagation of probability waves goes «non-linearly». At the beginning and at the end of the half-period, the distribution of velocities $\langle v \rangle_1^{1,0.1}$ changes rapidly, and in the middle of the half-period it is much slower (see Fig. 8). Due to this feature, the scales of the «time slices» are all different in Fig. 8.

For quantum states with numbers $\mu > 1$, distributions are obtained in the form of «superposition» of the distributions shown in Figs. 7, 8. An example of such a «superposition» for the state numbered $\mu = 5$ and $\beta = 0.1$ is shown in Fig. 9. The graphs shown in Fig. 9 were obtained for the same «time slices» as in Fig. 8. Note that according to expression (1.8), period $T_\mu \sim 1/\mu^2$, therefore, for $\mu = 5$ the oscillation frequency of the density of probabilities $f_{5,0.1}^1$ and $\overline{f}_{5,0.1}^1$ will be 25 times higher (see Fig. 9) than for functions $f_{1,0.1}^1$ and $\overline{f}_{1,0.1}^1$ ($\mu = 1$, see Figs. 7, 8).

The comparison of Figs. 8 and 9 shows that the distributions for the state numbered $\mu = 5$ (see Fig. 9) «consist» of the distributions for state numbered $\mu = 1$ (see Fig. 8). A similar picture is observed for the probability density distributions $f_{1,0.1}^1(x,0)$ (see Fig. 6) and $f_{5,0.1}^1(x,0)$ (see Fig. 3).

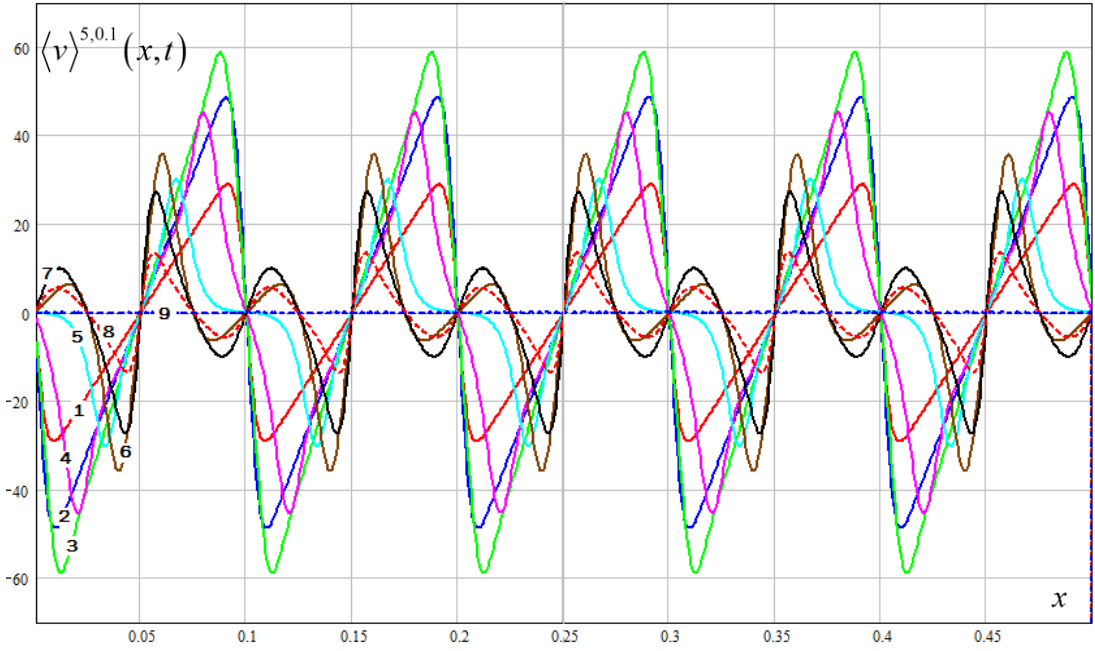

Fig. 9 Evolution of the distribution of velocity field $\langle v \rangle_1^{5,0.1}$ during half-period $t \in (0, T_1/2)$

With an increase of $\beta$, the evolutionary picture for distributions $f_{\mu,\beta}^1$ and $\langle v \rangle_1^{\mu,\beta}$, according to Theorems 3 and 7, degenerates into a stationary distribution.

### §3 The Gibbs distribution

Expression (1.12) may be analyzed in terms of averages over the Gibbs distribution. From definition $P_{n,k}^\mu$ (2.2) it follows that

$$(2k+1)^2 = \frac{\left[P_{k,k}^\mu\right]^2}{2m}\frac{1}{E_\mu}. \tag{3.1}$$

Using (3.1) we rewrite expression (1.12) and obtain

$$\bar{f}_{\mu,\beta}^1(x) = \frac{2}{N(\beta)}\sum_\kappa e^{-\frac{\pi\beta}{2}\frac{\hbar^2\kappa^2}{2m}\frac{1}{E_\mu}}\sin^2(\kappa x), \quad P_{k,k}^\mu = \hbar\kappa, \tag{3.2}$$

$$N(\beta) = l\sum_{k=-\infty}^{+\infty} e^{-\frac{\pi\beta}{2}(2k+1)^2} = l\sum_\kappa e^{-\frac{\pi\beta}{2}\frac{\hbar^2\kappa^2}{2m}\frac{1}{E_\mu}},$$

where $\kappa = \frac{\pi\mu}{l}(2k+1)-$ is a wave number ($\kappa \neq 0$). Let us introduce the notations

$$\mathcal{E}_\kappa \stackrel{\text{det}}{=} \frac{\hbar^2\kappa^2}{2m}, \quad \beta \stackrel{\text{det}}{=} \frac{\pi\beta}{2E_\mu} = \frac{1}{\tau}, \quad Z(\beta) \stackrel{\text{det}}{=} \sum_\kappa e^{-\beta\mathcal{E}_\kappa} = \frac{1}{l}N(\beta), \tag{3.3}$$

$$w_\kappa \stackrel{\text{det}}{=} \frac{1}{Z(\beta)} e^{-\beta \mathcal{E}_\kappa}, \qquad \sum_\kappa w_\kappa = 1,$$

where $\mathcal{E}_\kappa -$ is the kinetic energy corresponding to wave number $\kappa$, $\beta = \frac{1}{\tau}$ is an analogue of inverse temperature $\tau$, $Z(\beta) -$ is an analogue of statistical sum of the Gibbs distribution, $w_\kappa -$ is an analogue of the Gibbs distribution. With the use of notations (3.3), expression (3.2) will take the form:

$$\bar{f}^1_{\mu,\beta}(x) = \frac{2}{l} \frac{1}{Z(\beta)} \sum_\kappa e^{-\beta \mathcal{E}_\kappa} \sin^2(\kappa x) = \frac{2}{l} \left\langle \sin^2(\kappa x) \right\rangle_{Gibbs}. \qquad (3.4)$$

Thus, time-averaged (averaged over the period) time-dependent probability density $f^1_{\mu,\beta}$ is equal to the average by Gibbs of all wave packets (3.4). Let us analyze result (3.4). As the «temperature» of the quantum system decreases $\tau \to 0+$, the inverse temperature increases $\beta \to +\infty$, which, according to Theorems 3 and 5, leads to the «freezing» of the system, that is

$$\lim_{\beta \to +\infty} f^1_{\mu,\beta}(x,t) = \lim_{\beta \to +\infty} \bar{f}^1_{\mu,\beta}(x) = f^1_\mu(x), \qquad (3.5)$$

from this

$$\lim_{\beta \to +\infty} \left\langle \sin^2(\kappa x) \right\rangle_{Gibbs} = \sin^2\left( \frac{\sqrt{2mE_\mu}}{\hbar} x \right) = \sin^2\left( \frac{\pi \mu}{l} x \right).$$

In this case, the average velocity field $\langle v \rangle_1^{\mu,\beta}(x,t)$, in accordance with Theorem 7, also tends to zero

$$\lim_{\beta \to +\infty} \langle v \rangle_1^{\mu,\beta}(x,t) = 0. \qquad (3.6)$$

This result has been repeatedly discussed above and is illustrated in Figs. 2, 4. With the increasing temperature $\tau \to +\infty$, the inverse temperature decreases $\beta \to 0$, which leads to a strong dependency of the quantum system on time, i.e. the presence of rapidly oscillating probability density waves (see Figs. 1, 2, 8, 9).

Let us formulate the theorem.

**Theorem 8** *Energy $\mathcal{E}$ averaged by Gibbs (3.3) has the form*

$$\left\langle \mathcal{E}_\mu \right\rangle_{Gibbs}(\beta) = \sum_\kappa w_\kappa \mathcal{E}_\kappa = -\frac{d}{d\beta} \ln Z(\beta), \qquad (3.7)$$

$$Z(\beta) = \theta_1\left( -\frac{1}{2}, i\frac{4E_\mu}{\pi}\beta \right), \qquad (3.8)$$

*in this case, the passage to the limit is valid*

$$\lim_{\beta \to +\infty} \left\langle \mathcal{E}_\mu \right\rangle_{Gibbs}(\beta) = E_\mu. \qquad (3.9)$$

The proof of Theorem 8 is given in the Appendix.

Fig. 10 shows the dependence of average energy $\langle \mathcal{E}_\mu \rangle_{Gibbs}(\beta)$ on inverse temperature $\beta$ for various quantum states $\mu$. It is seen that, according to Theorem 8, at $\beta \to +\infty$ (cooling of the system) energy spectrum $\langle \mathcal{E}_\mu \rangle_{Gibbs}$ tends to the spectrum $E_\mu$ of time-independent solution (i.3).

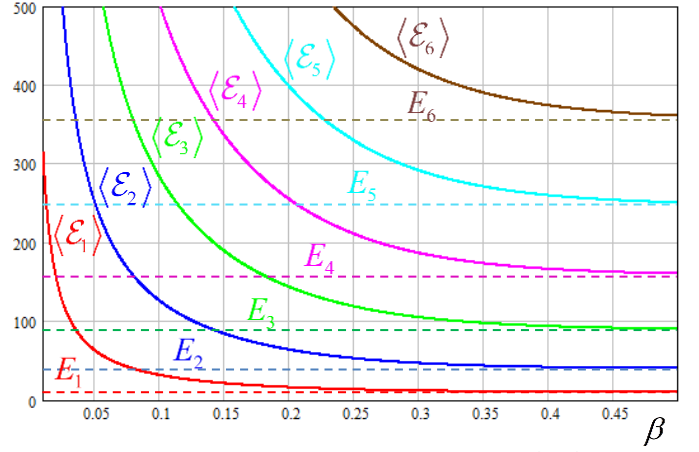

Fig. 10 Dependence of average energy $\langle \mathcal{E}_\mu \rangle_{Gibbs}$ on the inverse temperature

Using the Wigner function (2.2), we find the coordinate distribution of the average energy value (according to (i.1), the potential energy inside the well equals to zero):

$$\langle \mathcal{E}_{\mu,\beta} \rangle (x,t) = \frac{\hbar}{f_{\mu,\beta}^1(x,t)} \int_0^l \frac{p^2}{2m} W_{\mu,\beta}(x,p,t) dp, \qquad (3.10)$$

$$\langle \mathcal{E}_{\mu,\beta} \rangle (x,t) = \frac{E_\mu}{N(\beta) f_{\mu,\beta}^1(x,t)} \sum_{n,k=-\infty}^{+\infty} e^{-\frac{\pi\beta}{4}\left[(2n+1)^2 + (2k+1)^2\right]} (n+k+1)^2 T_{|k-n|}\left[\cos \vartheta_{n,k}^\mu(x,t)\right],$$

where expression $f_{\mu,\beta}^{1,2}(x,v,t) = \hbar W_{\mu,\beta}(x,p,t)$ is taken into consideration.

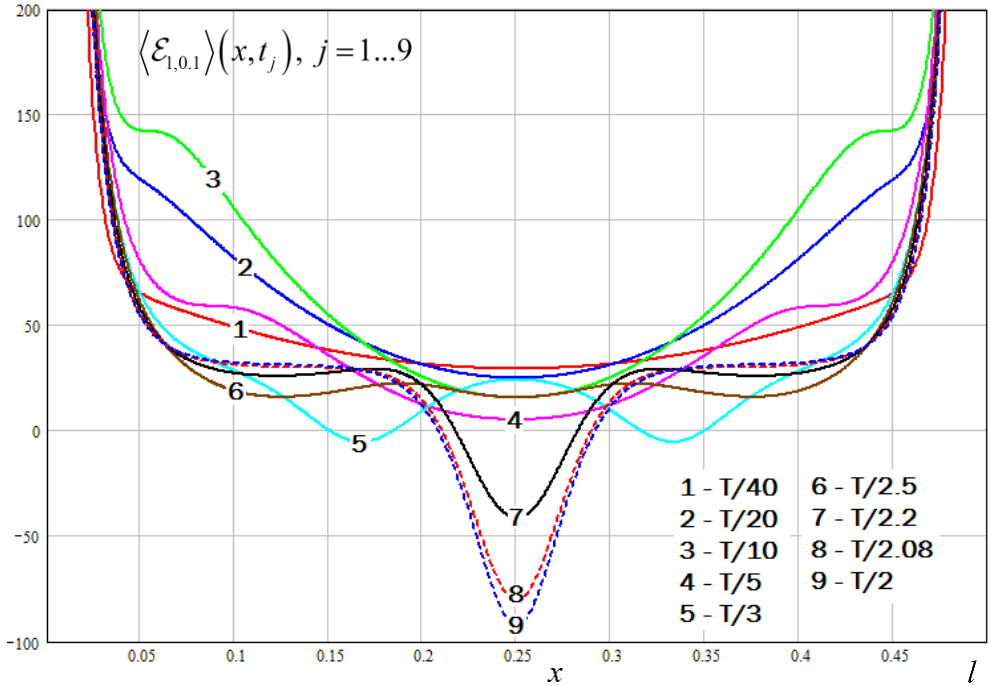

Fig. 11 Evolution of average energy $\langle \mathcal{E}_{1,0.1} \rangle$ inside the potential well

Distribution (3.10) is periodic with period $T_\mu$. Fig. 11 shows the evolution of the redistribution of energy $\langle \mathcal{E}_{\mu,\beta} \rangle (x,t)$ inside the potential well for the quantum state with number $\mu = 1$ and $\beta = 0.1$. A feature of the evolution of energy $\langle \mathcal{E}_{\mu,\beta} \rangle (x,t)$ is the presence of negative values (see Fig. 11, central domain, instants of time $t_5, t_7, t_8, t_9$), which are caused by negative values of the Wigner function (2.1). The negative values of energy $\langle \mathcal{E}_{\mu,\beta} \rangle$ are also associated with the definition of quantum potential $Q(x,t)$, which enters into the Hamilton-Jacobi equation [20] and is used in the «pilot-wave» theory by de Broglie-Bohm [21-24]

$$\frac{\partial \Phi_{\mu,\beta}}{\partial t} = -\frac{2}{\hbar} \langle \mathcal{E}_{\mu,\beta} \rangle = -\frac{2}{\hbar}\left[\frac{m}{2}\left|\langle v \rangle^{\mu,\beta}\right|^2 + e\chi_{\mu,\beta}\right], \qquad (3.11)$$

$$e\chi_{\mu,\beta} = U + Q_{\mu,\beta}, \qquad Q_{\mu,\beta} = -\frac{\hbar^2}{2m}\frac{1}{\sqrt{f_{\mu,\beta}^1}}\frac{\partial^2}{\partial x^2}\sqrt{f_{\mu,\beta}^1}, \qquad (3.12)$$

$$\Phi_{\mu,\beta} = 2\varphi_{\mu,\beta} + 2\pi k, \; k \in \mathbb{Z}, \quad \varphi_{\mu,\beta} = \frac{S_{\mu,\beta}}{\hbar}, \quad \langle v \rangle^{\mu,\beta} = \frac{\hbar}{2m}\frac{\partial \Phi_{\mu,\beta}}{\partial x}, \qquad (3.13)$$

where $\varphi_{\mu,\beta}(x,t)$ – is the phase of wave function $\Psi_{\mu,\beta}(x,t)$ (1.4); $S_{\mu,\beta}$ – is the action. Fig. 12 shows the distribution of the quantum potential for instants of time $t_5, t_7, t_8, t_9$, for which the distribution of energy $\langle \mathcal{E}_{\mu,\beta} \rangle$ in Fig. 11 has negative values domains. According to the Hamilton-Jacobi equation (3.11), energy $\langle \mathcal{E}_{\mu,\beta} \rangle$ is the sum of the kinetic energy (determined by expression (2.11), see Fig. 8) and potential energy, which completely (see expression (i.1)) coincides with quantum potential $Q_{\mu,\beta}$ (3.12) (see Fig. 12). Considering the above and comparing the distributions in Fig. 11 with distributions in Fig. 12 it is seen that the negative values of $\langle \mathcal{E}_{\mu,\beta} \rangle$ are associated with the negative values of quantum potential $Q_{\mu,\beta}$.

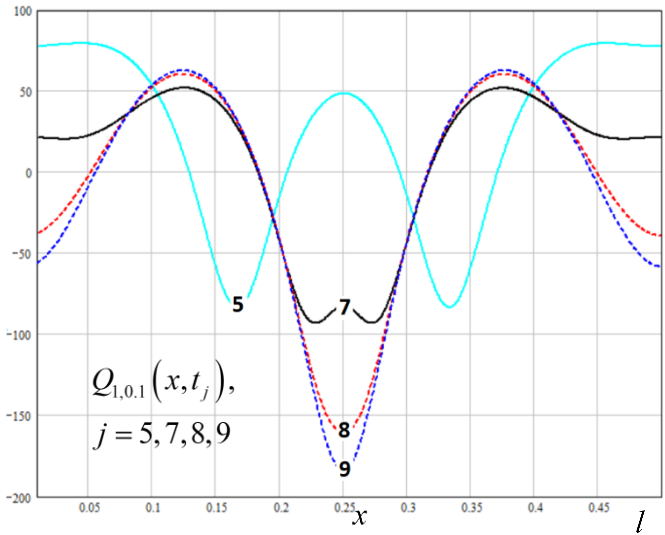

Fig. 12 Evolution of quantum potential $\langle Q_{1,0.1} \rangle$ inside the potential well

At the edges of the potential well, energy function $\langle \mathcal{E}_{\mu,\beta} \rangle$ has poles, since the denominator of expression (3.10) $f_{1,\beta}^1(x,t)$ is equal to zero, that is $f_{1,\beta}^1(0,t) = f_{1,\beta}^1(l,t) = 0$. For quantum states with number $\mu > 1$, the evolution of the distribution of energy $\langle \mathcal{E}_{\mu,\beta} \rangle (x,t)$ will have a structure that is similar to the described one but periodic in coordinate (the number of periods in the potential well will be equal to number of the quantum state $\mu$). A similar structure of distributions was observed when comparing Fig. 8 ($\mu = 1$) with Fig. 9 ($\mu = 5$). The number of poles of energy $\langle \mathcal{E}_{\mu,\beta} \rangle (x,t)$ will be equal to $\mu + 1$.

**Theorem 9** *Let the average values of $\langle \bar{\mathcal{E}}_{\mu,\beta} \rangle(x)$ and $\langle\langle \bar{\mathcal{E}}_{\mu,\beta} \rangle\rangle$ be defined as*

$$\langle \bar{\mathcal{E}}_{\mu,\beta} \rangle(x) \stackrel{\text{det}}{=} \frac{1}{\bar{f}^1_{\mu,\beta}(x)} \int_0^{T_\mu} f^1_{\mu,\beta}(x,t) \langle \mathcal{E}_{\mu,\beta} \rangle(x,t) dt, \qquad (3.14)$$

$$\langle\langle \bar{\mathcal{E}}_{\mu,\beta} \rangle\rangle \stackrel{\text{det}}{=} \frac{1}{\bar{f}^0_{\mu,\beta}} \int_0^l \bar{f}^1_{\mu,\beta}(x) \langle \bar{\mathcal{E}}_{\mu,\beta} \rangle(x) dx, \quad \bar{f}^0_{\mu,\beta} \stackrel{\text{det}}{=} \int_0^l \bar{f}^1_{\mu,\beta}(x) dx,$$

*then the following representations are valid*

$$\langle \bar{\mathcal{E}}_{\mu,\beta} \rangle(x) = \frac{1}{l} \frac{\langle \mathcal{E}_\mu \rangle_{Gibbs}(\beta)}{\bar{f}^1_{\mu,\beta}(x)} = \frac{\langle \mathcal{E}_\mu \rangle_{Gibbs}(\beta)}{2\langle \sin^2(\kappa x) \rangle_{Gibbs}}, \quad \langle\langle \bar{\mathcal{E}}_{\mu,\beta} \rangle\rangle = \langle \mathcal{E}_\mu \rangle_{Gibbs}(\beta), \qquad (3.15)$$

*wherein*

$$\lim_{\beta \to +\infty} \langle \bar{\mathcal{E}}_{\mu,\beta} \rangle(x) = \frac{1}{l} \frac{E_\mu}{f^1_\mu(x)}, \qquad \lim_{\beta \to +\infty} \langle\langle \bar{\mathcal{E}}_{\mu,\beta} \rangle\rangle = E_\mu. \qquad (3.16)$$

The proof of Theorem 9 is given in the Appendix.

Fig. 13 shows the distributions of average energy $\langle \bar{\mathcal{E}}_{2,\beta} \rangle$ for a quantum state with number $\mu = 2$ and different values of the inverse temperature. Distribution $\langle \bar{\mathcal{E}}_{2,\beta} \rangle$ has three poles (at points $x = 0$, $x = l$, $x = l/2$), which are determined by the three zeros of distribution function $\bar{f}^1_{2,\beta}(x)$. The poles «break» (create infinitely high energy barriers) the distribution of energy $\langle \bar{\mathcal{E}}_{2,\beta} \rangle$ inside the potential well into two symmetric distributions (see Fig. 13) within two identical domains of size $l/2$ [25].

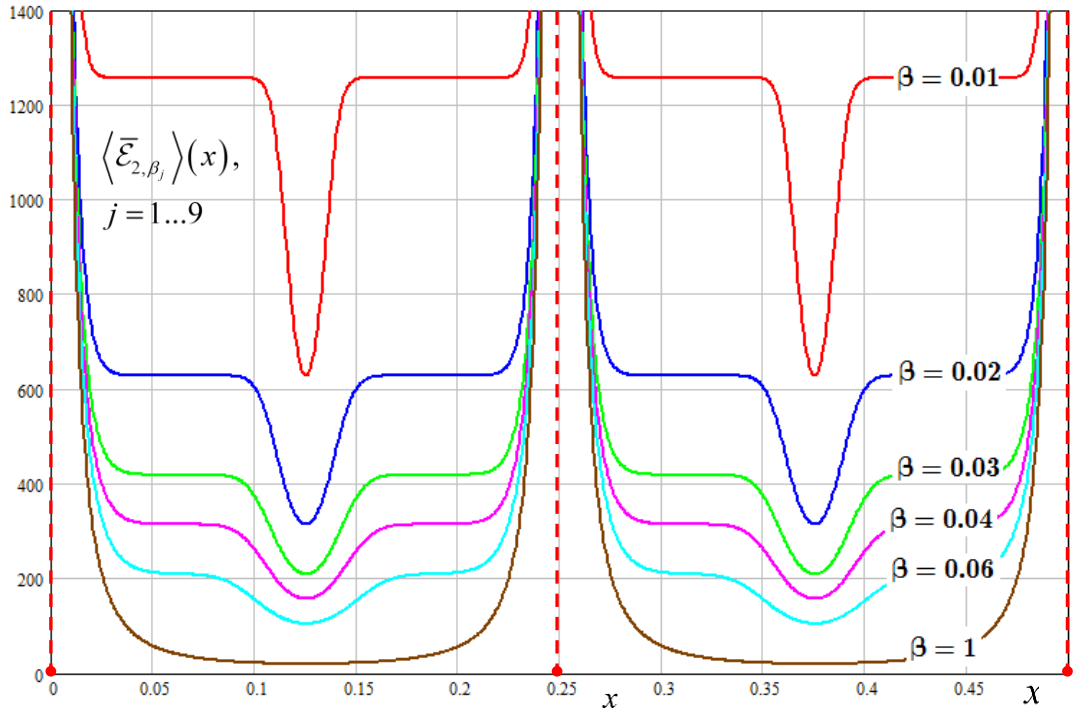

Fig. 13 Distribution of average energy $\langle \bar{\mathcal{E}}_{2,\beta} \rangle$ inside the potential well

In Fig. 13, it is seen that with a decrease of the «temperature» of the quantum system ( $\beta \to +\infty$ ) the energy inside each domain is equalized, and with an increase in the «temperature» ( $\beta \to 0$ ) the energy has a strong inhomogeneity of distribution in the centers of each domain (see Fig. 13). The comparison of the distributions of energy $\langle \bar{\mathcal{E}}_{2,\beta} \rangle$ in Fig. 13 with the distributions of probability density $\bar{f}^1_{\mu,\beta}(x)$, for instance, in Fig. 3, shows that all these distributions have strong inhomogeneity in the middles of the domains. This inhomogeneity is most pronounced at a high temperature ( $\beta \to 0$ ) and completely disappears at a low temperature ( $\beta \to +\infty$ ). The picture of this behavior becomes clearer if one takes into account the distribution of velocities $\langle v \rangle^{\mu,\beta}$ in the potential well (see Figs. 8, 9). For convenience of reasoning, let us consider the concept of a continuous medium corresponding to probability density $\bar{f}^1_{\mu,\beta}(x)$. At a low temperature, almost all «particles» of the continuous medium are motionless, since velocity $\langle v \rangle^{\mu,\beta} \to 0$ at $\beta \to +\infty$ (see Theorem 7, expression (2.14)). The energy of such particles is constant, which corresponds to distribution $\langle \bar{\mathcal{E}}_{2,\beta} \rangle$ in Fig. 13 ( $\beta = 1$ ). The particles at rest are distributed inside the potential domain (see Fig. 3) according to law $f^1_\mu(x)$. With increasing temperature, oscillating probability flows arise (see Fig. 8). During the first half of period $t \in (0, T_1/2)$, the main flows of particles of the medium diverge from the center of the domain in opposite directions. The particles of the medium located on the left side of the potential well move to the left, and the particles from the right side move to the right (see Fig. 8). In the middle of period $t = T_1/2$, the medium velocity becomes zero, after which, during the second half-period $t \in (T_1/2, T_1)$, the flows begin to move in opposite directions − towards each other. The particles of the medium, which are in the center of the domains («central particles»), have zero velocity (see Figs. 8, 9), since mutually compensating flows influence on them from the left and right. At low temperatures, a significant part of the «central particles» is at rest (or has low energy $\langle \bar{\mathcal{E}}_{2,\beta} \rangle$, see Fig. 13, $\beta = 0.06$ ) and oscillations are performed basically by the «peripheral particles» of the medium (having high energy $\langle \bar{\mathcal{E}}_{2,\beta} \rangle$, see Fig. 13, $\beta = 0.06$ ).

An increase in temperature (an increase in the velocity amplitude and energy $\langle \bar{\mathcal{E}}_{2,\beta} \rangle$) leads to an increase in the number of particles participating in the oscillation (see Fig. 13, $\beta = 0.01$) and it also leads to a decrease of the number of remaining «central» particles (having low energy $\langle \bar{\mathcal{E}}_{2,\beta} \rangle$, see Fig. 13, $\beta = 0.01$) not participating in oscillatory process (see Fig. 3). Indeed, Fig. 3 shows the formation of a horizontal plateau in the distribution function $\bar{f}^1_{\mu,\beta}(x)$ to the left and right of the center of the domain. In the center of the domain, there occurs narrowing of the width of the peak of distribution $\bar{f}^1_{\mu,\beta}(x)$, which indicates a decrease in the fixed number of «central particles». In the limiting case at $\beta \to 0$, function $\bar{f}^1_{\mu,\beta}(x)$ degenerates into a set of $\delta$ −functions with centers in the middle of the domains (the so-called «Dirac comb»). Thus, in the limit, only one stationary «central particle» remains (see Figs. 1, 3).

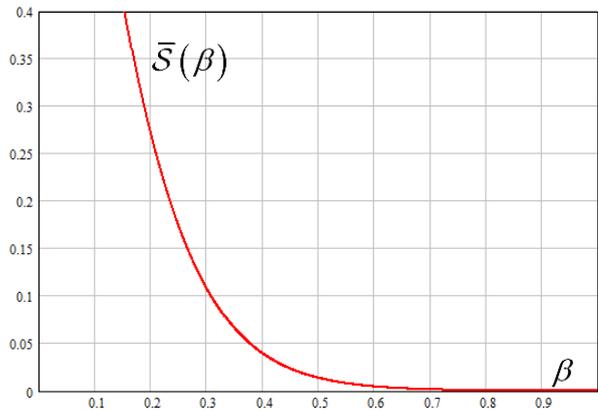

Fig. 14 Distribution of entropy $\bar{\mathcal{S}}(\beta)$

Let us consider the concept of thermodynamic entropy $\mathcal{S}$ of a quantum system. Taking into account the second law of thermodynamics, we get (3.3)

$$d\langle\mathcal{E}_\mu\rangle_{Gibbs} = \frac{\tau}{k_B} d\mathcal{S}, \quad d\mathcal{S} = k_B \beta \, d\langle\mathcal{E}_\mu\rangle_{Gibbs}, \qquad (3.17)$$

where $k_B$ — is the Boltzmann constant.

***Theorem 10*** *The thermodynamic entropy* $\mathcal{S}$ *(3.17) of a quantum system with potential (i.1) satisfies the following representation*

$$\mathcal{S}(\beta) = -k_B \ln \mathcal{Z}(\beta), \qquad (3.18)$$

$$\mathcal{Z}(\beta) = \frac{2}{Z(\beta)} e^{-\beta\langle\mathcal{E}_\mu\rangle_{Gibbs}(\beta)},$$

*wherein*

$$\lim_{\beta \to +\infty} \mathcal{S}(\beta) = 0, \quad \lim_{\beta \to 0+} \mathcal{S}(\beta) = +\infty. \qquad (3.19)$$

The proof of Theorem 10 is given in the Appendix.

**Remark**

Let us note that the entropy (3.18) does not depend on the number of the quantum state $\mu$. Indeed, we can rewrite the expression (3.18) in terms of the variable $\beta$ (3.3):

$$\mathcal{S}(\beta) = \bar{\mathcal{S}}(\beta) = -k_B \ln\left\{ \frac{2l}{N(\beta)} \exp\left[ -\frac{\pi l}{2} \frac{\beta}{N(\beta)} \sum_{k=-\infty}^{+\infty} (2k+1)^2 e^{-\frac{\pi\beta}{2}(2k+1)^2} \right] \right\}. \qquad (3.20)$$

Fig. 14 shows the dependence of entropy (3.20) on the inverse temperature. In accordance with Theorem 10, the entropy tends to zero with decreasing temperature («freezing») of the quantum system. As the temperature rises, the entropy grows indefinitely (see Fig. 14). Entropy $\bar{\mathcal{S}}(\beta)$ is a strictly monotonic function.

It can be concluded that the system under consideration obeys the three conservation laws [14, 19]:

***The probability / mass conservation law***

$$\frac{\partial f^1_{\mu,\beta}}{\partial t} + \frac{\partial}{\partial x}\left[ f^1_{\mu,\beta} \langle\vec{v}\rangle_1^{\mu,\beta} \right] = 0, \qquad (3.21)$$

where $f^1_{\mu,\beta}$ and $\langle\vec{v}\rangle_1^{\mu,\beta}$ are defined by expressions (1.6) and (2.11) respectively.

***The momentum conservation law***

$$\frac{d\langle v\rangle_1^{\mu,\beta}}{dt} = \left( \frac{\partial}{\partial t} + \langle v\rangle_1^{\mu,\beta} \frac{\partial}{\partial x} \right)\langle v\rangle_1^{\mu,\beta} = -\frac{1}{f^1_{\mu,\beta}} \frac{\partial P_{11}^{\mu,\beta}}{\partial x} + \langle\langle\dot{v}\rangle\rangle_1^{\mu,\beta}, \qquad (3.22)$$

where $\langle\langle \dot{v}\rangle\rangle_1^{\mu,\beta} = 0$ due to (2.8); $\mathrm{P}_{11}^{\mu,\beta} = \int_{-\infty}^{+\infty} f_{\mu,\beta}^{1,2}\left(v - \langle v\rangle_1^{\mu,\beta}\right)^2 dv$ is a pressure:

$$\mathrm{P}_{11}^{\mu,\beta} = \frac{1}{mN(\beta)} \sum_{n,k=-\infty}^{+\infty} e^{-\frac{\pi\beta}{4}\left[(2n+1)^2 + (2k+1)^2\right]} \left[P_{n,k}^{\mu} - m\langle v\rangle_1^{\mu,\beta}(x,t)\right]^2 T_{|k-n|}\left[\cos\vartheta_{n,k}^{\mu}(x,t)\right], \quad (3.23)$$

where (2.11) is taken into account for the Wigner function $f_{\mu,\beta}^{1,2}$ (2.2). Let us note that the pressure $\mathrm{P}_{11}^{\mu,\beta}$ is related with quantum potential $Q_{\mu,\beta}$ (see Fig. 12, (3.12)) [19, 27]:

$$-\frac{1}{f_{\mu,\beta}^1}\frac{\partial \mathrm{P}_{11}^{\mu,\beta}}{\partial x} = \frac{\hbar^2}{2m^2}\frac{\partial}{\partial x}\left(\frac{1}{f_{\mu,\beta}^1}\frac{\partial^2 \sqrt{f_{\mu,\beta}^1}}{\partial x^2}\right) = -\frac{1}{m}\frac{\partial Q_{\mu,\beta}}{\partial x}. \quad (3.24)$$

Taking (3.23) into account the expression (3.24) can be rewritten in an explicit form.

### *The energy conservation law*

$$\frac{\partial}{\partial t}\left[\frac{f_{\mu,\beta}^1}{2}\left|\langle v\rangle_1^{\mu,\beta}\right|^2 + \frac{1}{2}\mathrm{P}_{11}^{\mu,\beta}\right] + \frac{\partial}{\partial x}\left[\frac{f_{\mu,\beta}^1}{2}\left|\langle v\rangle_1^{\mu,\beta}\right|^2 \langle v\rangle_1^{\mu,\beta} + \frac{1}{2}\langle v\rangle_1^{\mu,\beta}\mathrm{P}_{11}^{\mu,\beta} + \langle v\rangle_1^{\mu,\beta}\mathrm{P}_{11}^{\mu,\beta} + \frac{1}{2}\mathrm{P}_{111}^{\mu,\beta}\right] =$$

$$= \int_{-\infty}^{+\infty} f_{\mu,\beta}^{1,2}\langle \dot{v}\rangle_1^{\mu,\beta} v\, dv = 0, \quad (3.25)$$

where the terms in equation (3.25) correspond to the following macroscopic quantities: $\frac{f_{\mu,\beta}^1}{2}\left|\langle v\rangle_1^{\mu,\beta}\right|$ is the kinetic energy density; $\frac{1}{2}\mathrm{P}_{11}^{\mu,\beta}$ is the internal energy density; $\frac{f_{\mu,\beta}^1}{2}\left|\langle v\rangle_1^{\mu,\beta}\right|^2\langle v\rangle_1^{\mu,\beta}$ is the kinetic energy flux density; $\frac{1}{2}\langle v\rangle_1^{\mu,\beta}\mathrm{P}_{11}^{\mu,\beta}$ – is the flow of internal energy; $\langle v\rangle_1^{\mu,\beta}\mathrm{P}_{11}^{\mu,\beta}$ characterizes the «work of probability»; $\frac{1}{2}\mathrm{P}_{111}^{\mu,\beta}$ – is the «heat» flux,

$$\mathrm{P}_{111}^{\mu,\beta} = \int_{-\infty}^{+\infty} f_{\mu,\beta}^{1,2}\left(v - \langle v\rangle_1^{\mu,\beta}\right)^3 dv$$

$$\mathrm{P}_{111}^{\mu,\beta} = \frac{1}{mN(\beta)} \sum_{n,k=-\infty}^{+\infty} e^{-\frac{\pi\beta}{4}\left[(2n+1)^2 + (2k+1)^2\right]} \left[P_{n,k}^{\mu} - m\langle v\rangle_1^{\mu,\beta}(x,t)\right]^3 T_{|k-n|}\left[\cos\vartheta_{n,k}^{\mu}(x,t)\right].$$

The right-hand side of equation (3.25) $\int_{-\infty}^{+\infty} f_{\mu,\beta}^{1,2}\langle \dot{v}\rangle_1^{\mu,\beta} v\, dv = 0$ is the mean value of the external forces (where (2.8) is taken into account).

Conservation laws (3.21), (3.22) and (3.25) upon «freezing» the quantum system ($\beta \to +\infty$) degenerate into trivial equations «0 = 0». The energy of a quantum system according to the Hamilton-Jacobi equation (3.11) has the following asymptotics

$$\langle \mathcal{E}_{\mu,\beta}\rangle = \frac{m}{2}\left|\langle v\rangle^{\mu,\beta}\right|^2 + Q_{\mu,\beta},$$

$$\lim_{\beta \to +\infty} \langle \mathcal{E}_{\mu,\beta} \rangle = \lim_{\beta \to +\infty} Q_{\mu,\beta} = Q_\mu = -\frac{\hbar^2}{2m} \frac{1}{\sqrt{f_\mu^1}} \frac{\partial^2}{\partial x^2} \sqrt{f_\mu^1} = E_\mu, \qquad (3.26)$$

where Theorems 3, 7 and representation (i.1), (i.4) are taken into account. Note that equality (3.26) coincides with expression (3.9). Thus, the energy of the «frozen» system coincides with the quantum potential $Q_\mu$.

**Conclusions**

The obtained time-dependent solution of the Schrödinger equation allows one to take a fresh look at the whole set of time-independent solutions in general. The following analogy can be applied here. The time-dependent heat-conduction equation under certain initial-boundary conditions gives solutions that converge over time to time-independent solutions corresponding to the thermodynamic equilibrium of the system.

The Schrödinger equation is mathematically similar to the heat-conduction equation and, as shown in this work, has a similar behavior. In the limiting case, as the «quantum system cools down», the time-dependent solution converges to a time-independent solution. Despite the known weirdness of the quantum world, dynamic processes in such a system have a clear interpretation in terms of classical physics, that is, they give a chance to «understand» the mechanisms of quantum mechanics.

**Acknowledgements**

This research has been supported by the Interdisciplinary Scientific and Educational School of Moscow University «Photonic and Quantum Technologies. Digital Medicine».

**Appendix**

*Proof of Theorem 1*

Let us consider a univariate $\theta$-function with characteristics $a$ and $b$:

$$\theta[a,b](z,\tau) = \sum_{k=-\infty}^{+\infty} e^{\pi i \tau (k+a)^2 + 2\pi i (z+b)(k+a)}, \qquad (A.1)$$

where $z \in \mathbb{C}$, $\tau = \alpha + i\beta$, $\beta > 0$. The condition $\beta > 0$ is necessary for the convergence of series (A.1). Differentiating $\theta$-function (A.1) over variables $z$ and $\tau$, one may show that the heat-conduction equation is valid:

$$\frac{\partial^2 \theta[a,b](z,\tau)}{\partial z^2} = 4\pi i \frac{\partial \theta[a,b](z,\tau)}{\partial \tau}, \qquad (A.2)$$

which, when changing the variables,

$$z = \frac{\sqrt{2m\varepsilon}}{\hbar} x, \quad \tau = -\frac{4\pi\varepsilon}{\hbar} t + i\beta, \qquad (A.3)$$

takes the form

$$i\hbar \frac{\partial \Theta(x,t)}{\partial t} = -\frac{\hbar^2}{2m} \frac{\partial^2 \Theta(x,t)}{\partial x^2}, \tag{A.4}$$

$$\Theta(x,t) \stackrel{det}{=} \theta[a,b]\left(\frac{\sqrt{2m\varepsilon}}{\hbar}x, -\frac{4\pi\varepsilon}{\hbar}t + i\beta\right), \tag{A.5}$$

where $\varepsilon -$ is constant value. Equation (A.4) coincides with Schrödinger equation (i.2). Boundary conditions (i.2) may be met in the consideration of the particular case of $\theta$-function (A.5) with characteristics $a = b = \frac{1}{2}$, which correspond to $\theta_1 -$ function:

$$\theta_1(z,\tau) \stackrel{det}{=} -\theta\left[\frac{1}{2},\frac{1}{2}\right](z,\tau) = 0 \Leftrightarrow z = \mu + \nu\tau, \quad \mu,\nu \in \mathbb{Z}, \tag{A.6}$$

choosing $\nu = 0$, we obtain

$$\mu = \frac{\sqrt{2m\varepsilon_\mu}}{\hbar} l \Rightarrow \varepsilon_\mu = \frac{\hbar^2 \mu^2}{2ml^2}, \quad \tau = -\frac{4\pi\varepsilon_\mu}{\hbar}t + i\beta = -\frac{2\pi\hbar\mu^2}{ml^2}t + i\beta, \tag{A.7}$$

where $E_\mu = \pi^2 \varepsilon_\mu$. As a result, solution (A.5) will take the form:

$$\Theta_{\mu,\beta}(x,t) = \theta_1\left(\frac{\sqrt{2m\varepsilon_\mu}}{\hbar}x, -\frac{4\pi\varepsilon_\mu}{\hbar}t + i\beta\right) = \sum_{k=-\infty}^{+\infty} e^{\pi i\left(-\frac{4\pi\varepsilon_\mu}{\hbar}t + i\beta\right)\left(k+\frac{1}{2}\right)^2 + 2\pi i\left(\frac{\sqrt{2m\varepsilon_\mu}}{\hbar}x + \frac{1}{2}\right)\left(k+\frac{1}{2}\right)}, \tag{A.8}$$

$$\Theta_{\mu,\beta}(0,t) = \Theta_{\mu,\beta}(l,t) = 0,$$

where parameter $\beta > 0$ may vary. Let us fulfill the normalization condition for function (A.8).

$$N = \int_0^l \left|\Theta_{\mu,\beta}(x,t)\right|^2 dx. \tag{A.9}$$

We calculate value $\left|\Theta_{\mu,\beta}\right|^2 = \Theta_{\mu,\beta}\Theta^*_{\mu,\beta}$ and obtain

$$\left|\Theta^\beta_{\mu,\beta}\right|^2 = \sum_{n,k=-\infty}^{+\infty} e^{-i\frac{4\pi^2\varepsilon_\mu}{\hbar}t\left(k+\frac{1}{2}\right)^2 -\pi\beta\left(k+\frac{1}{2}\right)^2 + 2\pi i\left(\frac{\sqrt{2m\varepsilon_\mu}}{\hbar}x + \frac{1}{2}\right)\left(k+\frac{1}{2}\right) + i\frac{4\pi^2\varepsilon_\mu}{\hbar}t\left(n+\frac{1}{2}\right)^2 -\pi\beta\left(n+\frac{1}{2}\right)^2 - 2\pi i\left(\frac{\sqrt{2m\varepsilon_\mu}}{\hbar}x + \frac{1}{2}\right)\left(n+\frac{1}{2}\right)} =$$

$$= \sum_{n,k=-\infty}^{+\infty} e^{-\pi\beta\left[\left(k+\frac{1}{2}\right)^2 + \left(n+\frac{1}{2}\right)^2\right]} e^{i\frac{4\pi^2\varepsilon_\mu}{\hbar}\left[\left(n+\frac{1}{2}\right)^2 - \left(k+\frac{1}{2}\right)^2\right]t + 2\pi i\left(\frac{\sqrt{2m\varepsilon_\mu}}{\hbar}x + \frac{1}{2}\right)(k-n)},$$

$$\left|\Theta_{\mu,\beta}\right|^2 = \sum_{n,k=-\infty}^{+\infty} e^{-\pi\beta\left[\left(k+\frac{1}{2}\right)^2 + \left(n+\frac{1}{2}\right)^2\right]} e^{i2\pi\left[-\frac{2\pi\varepsilon_\mu}{\hbar}(n+k+1)t + \frac{\sqrt{2m\varepsilon_\mu}}{\hbar}x + \frac{1}{2}\right](k-n)}. \tag{A.10}$$

We substitute expression (A.10) into integral (A.9) and perform the change of the variables

$$\bar{x} = -\frac{2\pi\varepsilon_\mu}{\hbar}(n+k+1)t + \frac{1}{2} + \frac{\sqrt{2m\varepsilon_\mu}}{\hbar}x, \; d\bar{x} = \frac{\sqrt{2m\varepsilon_\mu}}{\hbar}dx, \tag{A.11}$$

$$\bar{x}_1 = -\frac{2\pi\varepsilon_\mu}{\hbar}(n+k+1)t + \frac{1}{2},$$

$$\bar{x}_2 = -\frac{2\pi\varepsilon_\mu}{\hbar}(n+k+1)t + \frac{1}{2} + \frac{\sqrt{2m\varepsilon_\mu}}{\hbar}l = -\frac{2\pi\varepsilon_\mu}{\hbar}(n+k+1)t + \frac{3}{2} = \bar{x}_1 + 1,$$

$$N = \int_0^l |\Theta_{\mu,\beta}(x,t)|^2 dx = \sum_{n,k=-\infty}^{+\infty} e^{-\pi\beta\left[\left(k+\frac{1}{2}\right)^2 + \left(n+\frac{1}{2}\right)^2\right]} I_{n,k}, \tag{A.12}$$

$$I_{n,k} = \int_0^l e^{i2\pi\left[-\frac{2\pi\varepsilon_\mu}{\hbar}(n+k+1)t + \frac{1}{2} + \frac{\sqrt{2m\varepsilon_\mu}}{\hbar}x\right](k-n)} dx = l\int_{\bar{x}_1}^{\bar{x}_2} e^{i2\pi(k-n)\bar{x}} d\bar{x}.$$

If $n = k$, then $I_{n,n} = l(\bar{x}_2 - \bar{x}_1) = l$. At $n \neq k$

$$I_{n,k} = \frac{le^{i2\pi(k-n)\bar{x}}}{i2\pi(k-n)}\Bigg|_{\bar{x}_1}^{\bar{x}_1+1} = \frac{l}{i2\pi(k-n)}\left[e^{i2\pi(k-n)\bar{x}_1 + i2\pi(k-n)} - e^{i2\pi(k-n)\bar{x}_1}\right] =$$

$$= \frac{l}{i2\pi(k-n)}\left[e^{i2\pi(k-n)\bar{x}_1} - e^{i2\pi(k-n)\bar{x}_1}\right] = 0.$$

Thus,

$$\int_0^l |\Theta_{\mu,\beta}(x,t)|^2 dx = l\sum_{k=-\infty}^{+\infty} e^{-2\pi\beta\left(k+\frac{1}{2}\right)^2} = N(\beta), \tag{A.13}$$

That is the normalizing $N(\beta)$ does not depend on time $t$. Consequently, the solution of the Schrödinger equation may be represented in the form

$$\Psi_{\mu,\beta}(x,t) = \frac{1}{\sqrt{N(\beta)}}\theta_1\left(\mu\frac{x}{l}, -\mu^2\frac{2\pi\hbar}{ml^2}t + i\beta\right),$$

$$\Psi_{\mu,\beta}(0,t) = \Psi_{\mu,\beta}(l,t) = 0, \; \int_0^l |\Psi_{\mu,\beta}(x,t)|^2 dx = 1.$$

Theorem 1 is proved.

*Proof of Theorem 2*

Let us find expression $f_{\mu,\beta}^1(x,t)$.

$$f_{\mu,\beta}^1(x,t) = |\Psi_{\mu,\beta}(x,t)|^2 = \frac{1}{N(\beta)}\sum_{n,k=-\infty}^{+\infty} e^{-\frac{\pi\beta}{4}\left[(2k+1)^2 + (2n+1)^2\right]} e^{i\left[\pi\left(2\mu\frac{x}{l}+1\right) - \frac{E_\mu}{\hbar}4(n+k+1)t\right](k-n)}. \tag{A.14}$$

Taking into account that the real part in sum (A.14) is symmetric with respect to the indices $n$ and $k$, and the imaginary part (A.14) is antisymmetric, we obtain

$$f^1_{\mu,\beta}(x,t) = \frac{1}{N(\beta)} \sum_{n,k=-\infty}^{+\infty} e^{-\frac{\pi\beta}{4}\left[(2k+1)^2+(2n+1)^2\right]} \cos\left\{\left[\pi\left(2\mu\frac{x}{l}+1\right)-\frac{4E_\mu}{\hbar}(n+k+1)t\right](k-n)\right\}. \quad \text{(A.15)}$$

Probability density function (A.15) is periodic, therefore we find its period T

$$f^1_{\mu,\beta}(x,0) = f^1_{\mu,\beta}(x,T),$$

from here it follows that

$$\cos\left\{\left[\pi\left(2\mu\frac{x}{l}+1\right)-\frac{4E_\mu}{\hbar}(n+k+1)T\right](k-n)\right\} - \cos\left\{\pi\left(2\mu\frac{x}{l}+1\right)(k-n)\right\} = 0$$

$$\sin\left\{\left[\pi\left(2\mu\frac{x}{l}+1\right)-\frac{4E_\mu}{\hbar}(n+k+1)T\right](k-n)\right\}\sin\left\{\frac{2E_\mu}{\hbar}(n+k+1)(k-n)T\right\} = 0$$

$$\frac{2E_\mu}{\hbar}(n+k+1)(k-n)T = \pi j \Rightarrow \frac{2E_\mu}{\hbar}\lambda T = \pi j, \; j,\lambda \in \mathbb{Z},$$

$$T_\mu = \frac{\pi\hbar}{4E_\mu} = \frac{ml^2}{2\pi\hbar\mu^2}, \quad \text{(A.16)}$$

It is taken into consideration here that the minimum value of $\lambda$ giving a nonzero period T is $\lambda = 2$, for instance, at $k=1$, $n=0$.

The expression for density $f^1_{\mu,\beta}(x,t)$ may be represented in terms of the Chebyshev polynomials $T_{|k-n|}$, indeed

$$\cos\left\{\pi\left[\left(2\mu\frac{x}{l}+1\right)-\frac{t}{T_\mu}(n+k+1)\right](k-n)\right\} = \cos\left[\vartheta^\mu_{n,k}(x,t)|k-n|\right] = T_{|k-n|}\left(\cos\left[\vartheta^\mu_{n,k}(x,t)\right]\right),$$

$$\vartheta^\mu_{n,k}(x,t) \stackrel{\text{det}}{=} \pi\left(2\mu\frac{x}{l}+1\right) - \frac{\pi t}{T_\mu}(n+k+1), \quad \text{(A.17)}$$

$$f^1_{\mu,\beta}(x,t) = \frac{1}{N(\beta)} \sum_{n,k=-\infty}^{+\infty} e^{-\frac{\pi\beta}{4}\left[(2k+1)^2+(2n+1)^2\right]} T_{|k-n|}\left(\cos\left[\vartheta^\mu_{n,k}(x,t)\right]\right). \quad \text{(A.18)}$$

Theorem 2 is proved.

*Proof of Theorem 3*

Exponent (1.6) contains function $\eta(n,k) = (2n+1)^2 + (2k+1)^2$ that defines a paraboloid of revolution with a center (minimum) at point $n = k = -1/2$. Since $n, k \in \mathbb{Z}$, then there are only two mutually orthogonal straight lines passing through point $n = k = -1/2$, which integral-valued minima of function $\eta(n,k)$ lie on (see Fig. 15) $n - k = 0$ or $n + k + 1 = 0$. The summands of series (1.6) with indices $n = k$ are independent of variables $(x,t)$ and make contribution $\frac{2}{N(\beta)}e^{-\frac{\pi\beta}{2}}$ ($k=0$, $k=-1$, see Fig. 15). Dependence $n+k+1=0$ makes the main contribution $\frac{2}{N(\beta)}e^{-\frac{\pi\beta}{2}}\cos\left[\pi\left(2\mu\frac{x}{l}+1\right)\right]$ at $n=-1$, $k=0$ and $n=0$, $k=-1$. The denominator of expression (1.6) contains normalizing $N(\beta)$, which may be estimated (1.2) by value (see Fig.15) $N(\beta) \approx 2le^{-\frac{\pi\beta}{2}}$ at $k=0$ and $k=-1$. As a result, the following estimation is valid for expression (1.6) at large values of $\beta \gg 1$

$$f_{\mu,\beta}^1(x,t) \approx \frac{2}{N(\beta)} e^{-\frac{\pi\beta}{2}} \left[1 - \cos\left(2\pi\mu\frac{x}{l}\right)\right] \approx$$

$$\approx \frac{2}{l} \sin^2\left(\frac{\pi\mu x}{l}\right) = f_\mu^1(x). \qquad (A.19)$$

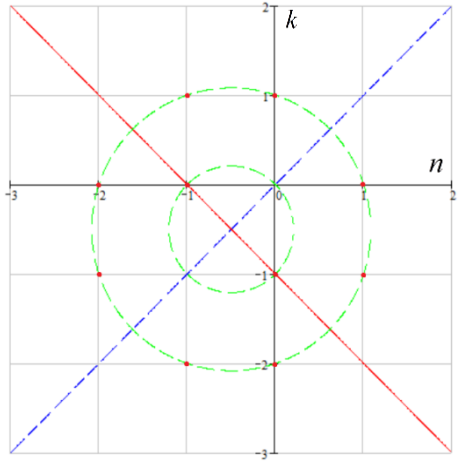

Fig. 15 Distribution of the minimums of function $\eta(n,k)$

Theorem 3 is proved.

*Proof of Theorem 4*

Let us perform direct calculations

$$\bar{f}_{\mu,\beta}^1 = \frac{1}{N(\beta)T_\mu} \sum_{n,k=-\infty}^{+\infty} e^{-\frac{\pi\beta}{4}\left[(2k+1)^2+(2n+1)^2\right]} \times$$

$$\times \int_0^{T_\mu} \cos\left\{\pi\left[\left(2\mu\frac{x}{l}+1\right) - \frac{t}{T_\mu}(n+k+1)\right](k-n)\right\} dt. \qquad (A.20)$$

The subintegral function in expression (A.20) may be represented in the form:

$$\cos\left\{\pi\left[\left(2\mu\frac{x}{l}+1\right)-\frac{t}{T_\mu}(n+k+1)\right](k-n)\right\} = \cos\left\{\pi\left(2\mu\frac{x}{l}+1\right)(k-n)\right\}\cos\left\{\frac{\pi t}{T_\mu}(n+k+1)(k-n)\right\} +$$

$$+\sin\left\{\pi\left(2\mu\frac{x}{l}+1\right)(k-n)\right\}\sin\left\{\frac{\pi t}{T_\mu}(n+k+1)(k-n)\right\}. \qquad (A.21)$$

The integral of periodic function (A.21) over its period is zero, provided that the argument of the function does not vanish, that is $(n+k+1)(k-n) \neq 0$. Therefore, for expression (A.21) it is necessary to consider the case $(n+k+1)(k-n) = 0$. For sin function in expression (A.21), the zero value of the argument will lead to the zero value of its integral, therefore, expression (A.20) will take the form:

$$\bar{f}_{\mu,\beta}^1 = \frac{1}{N(\beta)} \sum_{n,k=-\infty}^{+\infty} e^{-\frac{\pi\beta}{4}\left[(2k+1)^2+(2n+1)^2\right]} J_{n,k}^\mu(x), \qquad (A.22)$$

$$J_{n,k}^\mu(x) = \frac{1}{T_\mu} \cos\left[\pi\left(2\mu\frac{x}{l}+1\right)(k-n)\right] \int_0^{T_\mu} \cos\left[\frac{\pi t}{T_\mu}(n+k+1)(k-n)\right] dt, \qquad (A.23)$$

where

$$J_{n,k}^{\mu}(x) = \begin{cases} 1, & \text{if } k = n, \\ \cos\left[\pi\left(2\mu\dfrac{x}{l}+1\right)(2k+1)\right], & \text{if } k \ne n,\ n+k+1 = 0, \\ 0, & \text{if } (k-n)(n+k+1) \ne 0. \end{cases}$$

Substituting the values of integral (A.23) into expression (A.22) and taking into account (1.2), we obtain

$$\overline{f}_{\mu,\beta}^{1} = \frac{1}{N(\beta)}\sum_{k=-\infty}^{+\infty} e^{-\frac{\pi\beta}{2}(2k+1)^2} + \frac{1}{N(\beta)}\sum_{k=-\infty}^{+\infty} e^{-\frac{\pi\beta}{2}(2k+1)^2}\cos\left[\pi\left(2\mu\frac{x}{l}+1\right)(2k+1)\right] =$$

$$= \frac{1}{l} - \frac{1}{N(\beta)}\sum_{k=-\infty}^{+\infty} e^{-\frac{\pi\beta}{2}(2k+1)^2}\cos\left[2\pi\mu\frac{x}{l}(2k+1)\right] = \frac{1}{l} - \frac{1}{N(\beta)}\sum_{k=-\infty}^{+\infty} e^{-\frac{\pi\beta}{2}(2k+1)^2} +$$

$$+ \frac{2}{N(\beta)}\sum_{k=-\infty}^{+\infty} e^{-\frac{\pi\beta}{2}(2k+1)^2}\sin^2\left[\pi\mu\frac{x}{l}(2k+1)\right],$$

$$\overline{f}_{\mu,\beta}^{1} = \frac{2}{N(\beta)}\sum_{k=-\infty}^{+\infty} e^{-\frac{\pi\beta}{2}(2k+1)^2}\sin^2\left[(2k+1)\frac{\pi\mu}{l}x\right],$$

which was to be proved.

*Proof of Theorem 6*

Substituting expression (1.1) into integral (2.1), we obtain

$$\Psi_{\mu,\beta}^{*}\left(x-\frac{s}{2},t\right)\Psi_{\mu,\beta}\left(x+\frac{s}{2},t\right) = \frac{1}{N(\beta)}\sum_{n,k=-\infty}^{+\infty} e^{\left(-i\frac{\pi^2\varepsilon_\mu}{\hbar}t-\frac{\pi}{4}\beta\right)(2k+1)^2 + \left(i\frac{\pi^2\varepsilon_\mu}{\hbar}t-\frac{\pi}{4}\beta\right)(2n+1)^2} \times$$

$$\times e^{-\pi i(2n+1)\left(\frac{\sqrt{2m\varepsilon_\mu}}{\hbar}\left(x-\frac{s}{2}\right)+\frac{1}{2}\right) + \pi i(2k+1)\left(\frac{\sqrt{2m\varepsilon_\mu}}{\hbar}\left(x+\frac{s}{2}\right)+\frac{1}{2}\right)} =$$

$$= \frac{1}{N(\beta)}\sum_{n,k=-\infty}^{+\infty} e^{\left(i\frac{\pi^2\varepsilon_\mu}{\hbar}t-\frac{\pi}{4}\beta\right)(2n+1)^2 - \left(i\frac{\pi^2\varepsilon_\mu}{\hbar}t+\frac{\pi}{4}\beta\right)(2k+1)^2} e^{\pi i\left(\frac{2\sqrt{2m\varepsilon_\mu}}{\hbar}x+1\right)(k-n)} e^{\pi i\frac{\sqrt{2m\varepsilon_\mu}}{\hbar}(n+k+1)s},$$

$$W_{\mu,\beta}(x,p,t) = \frac{1}{2\pi\hbar N(\beta)}\sum_{n,k=-\infty}^{+\infty} e^{\left(i\frac{\pi^2\varepsilon_\mu}{\hbar}t-\frac{\pi}{4}\beta\right)(2n+1)^2 - \left(i\frac{\pi^2\varepsilon_\mu}{\hbar}t+\frac{\pi}{4}\beta\right)(2k+1)^2} e^{\pi i\left(\frac{2\sqrt{2m\varepsilon_\mu}}{\hbar}x+1\right)(k-n)} G_{n,k}^{\mu}(p), \quad \text{(A.24)}$$

$$G_{n,k}^{\mu}(\overline{p}) \stackrel{\text{det}}{=} \int_{-\infty}^{+\infty} e^{\frac{i\overline{p}s}{\hbar}}ds = 2\pi\delta(\overline{p}), \quad \overline{p} \stackrel{\text{det}}{=} \pi\sqrt{2m\varepsilon_\mu}(n+k+1) - p.$$

Let us transform the exponent in expression (A.24), we obtain

$$e^{-\frac{\pi\beta}{4}\left[(2n+1)^2 + (2k+1)^2\right]} e^{i\frac{\pi^2\varepsilon_\mu}{\hbar}\left[(2n+1)^2 - (2k+1)^2\right]t} e^{i\pi(k-n)\left(\frac{2\sqrt{2m\varepsilon_\mu}}{\hbar}x+1\right)} =$$

$$= e^{-\frac{\pi\beta}{4}\left[(2n+1)^2 + (2k+1)^2\right]}\left[\cos(\delta_1+\delta_2) + i\sin(\delta_1+\delta_2)\right],$$

(A.25)

where

$$\delta_1 = \frac{E_\mu}{\hbar}\left[(2n+1)^2 - (2k+1)^2\right]t, \qquad \delta_2 = \pi(k-n)\left(2\mu\frac{x}{l}+1\right), \qquad (A.26)$$

$$\delta_1 + \delta_2 = (k-n)\left\{\frac{2\pi\mu}{l}\left[-\frac{\mu\pi\hbar}{ml}(n+k+1)t + x\right] + \pi\right\}.$$

Function of sin in expression (A.25) is odd in indices $n, k$, consequently, when summing (A.24), only even terms with function cos will remain. Considering this and substituting (A.25) into expression (A.24), we obtain

$$W_{\mu,\beta}(x,p,t) = \frac{1}{\hbar N(\beta)} \sum_{n,k=-\infty}^{+\infty} (-1)^{k-n} e^{-\frac{\pi\beta}{4}\left[(2n+1)^2+(2k+1)^2\right]} \delta\left(\sqrt{2mE_\mu}(n+k+1)-p\right) \times$$
$$\times \cos\left\{(k-n)\frac{2\pi\mu}{l}\left[-\frac{\mu\pi\hbar}{ml}(n+k+1)t + x\right]\right\}. \qquad (A.27)$$

Let us introduce the notation for the momentum $P_{n,k}^\mu \stackrel{\text{det}}{=} \sqrt{2mE_\mu}(n+k+1)$, then the expression for the Wigner function will take the form

$$W_{\mu,\beta}(x,p,t) = \frac{1}{\hbar N(\beta)} \sum_{n,k=-\infty}^{+\infty} (-1)^{k-n} e^{-\frac{\pi\beta}{4}\left[(2n+1)^2+(2k+1)^2\right]} \delta\left(P_{n,k}^\mu - p\right) \cos\left[(k-n)\frac{2\pi\mu}{l}\left(-\frac{P_{n,k}^\mu}{m}t + x\right)\right],$$

or

$$W_{\mu,\beta}(x,p,t) = \frac{1}{\hbar N(\beta)} \sum_{n,k=-\infty}^{+\infty} e^{-\frac{\pi\beta}{4}\left[(2n+1)^2+(2k+1)^2\right]} \delta\left(P_{n,k}^\mu - p\right) T_{|k-n|}\left[\cos \vartheta_{n,k}^\mu(x,t)\right],$$

Since $\vartheta_{n,k}^\mu(x,t) = \frac{2\pi\mu}{l}\left(-\frac{P_{n,k}^\mu}{m}t + x\right) + \pi.$ Theorem 6 is proved.

*Proof of Theorem 7*

Substituting explicit expression (1.6) for function $f_{\mu,\beta}^1(x,t)$ into equation (2.5), we obtain the equation for average velocity $\langle v \rangle_1^{\mu,\beta}$. In order to simplify further calculations, we will use $\langle v \rangle$ instead of notation $\langle v \rangle_1^{\mu,\beta}$. The general solution $\langle v \rangle_{o.н}$ of inhomogeneous equation (2.4) / (2.5) may be represented as:

$$\langle v \rangle_{g.i.} = \langle v \rangle_{g.h.} + \langle v \rangle_{p.i.}, \qquad (A.28)$$

where $\langle v \rangle_{g.h.}$ – is the general solution to the homogeneous equation, $\langle v \rangle_{p.i.}$ – is the particular solution to the inhomogeneous equation. Let us find a solution to the homogeneous equation

$$\frac{\partial \langle v \rangle_1^{\mu,\beta}}{\partial x} + \frac{\partial S_{\mu,\beta}^1}{\partial x} \langle v \rangle_1^{\mu,\beta} = 0 \;\Rightarrow\; \ln\langle v \rangle_1^{\mu,\beta} = -\int \frac{\partial S_{\mu,\beta}^1}{\partial x} dx = -S_{\mu,\beta}^1 + const,$$

$$\ln\langle v \rangle_1^{\mu,\beta} = -\ln f_{\mu,\beta}^1 + const,$$

$$\langle v \rangle_{g.h.}(x,t) = \frac{C}{f^1_{\mu,\beta}(x,t)}, \tag{A.29}$$

To find $\langle v \rangle_{p.i.}$, let us use the method of variation of arbitrary constant $C$:

$$\frac{\partial f^1_{\mu,\beta}}{\partial t} + \frac{C}{f^1_{\mu,\beta}} \frac{\partial f^1_{\mu,\beta}}{\partial x} + \frac{\partial C}{\partial x} - \frac{C}{f^1_{\mu,\beta}} \frac{\partial f^1_{\mu,\beta}}{\partial x} = 0,$$

$$\frac{\partial C}{\partial x} = -\frac{\partial f^1_{\mu,\beta}}{\partial t}. \tag{A.30}$$

Substituting expression (1.6) into equation (A.30), we obtain

$$C(x,t) = -\int \frac{\partial f^1_{\mu,\beta}}{\partial t} dx = -\frac{\pi}{N(\beta)T_\mu} \sum_{n,k=-\infty}^{+\infty} e^{-\frac{\pi\beta}{4}\left[(2k+1)^2 + (2n+1)^2\right]}(n+k+1)(k-n)\Lambda^\mu_{n,k}, \tag{A.31}$$

we calculate the integral

$$\Lambda^\mu_{n,k} = \int \sin\left[(k-n)\vartheta^\mu_{n,k}\right]dx = \int \sin(Ax+B)dx,$$

where

$$(k-n)\vartheta^\mu_{n,k} = 2\mu\pi(k-n)\frac{x}{l} + \pi(k-n)\left[1 - \frac{t}{T_\mu}(n+k+1)\right] = Ax + B,$$

$$\bar{x} = Ax + B, \quad d\bar{x} = Adx,$$

$$\Lambda^\mu_{n,k} = \frac{1}{A}\int \sin(\bar{x})d\bar{x} = -\frac{1}{A}\cos(Ax+B) + const. \tag{A.32}$$

Substituting integral (A.32) into expression (A.31), we obtain

$$C(x,t) = \frac{l}{2\mu T_\mu N(\beta)} \sum_{n,k=-\infty}^{+\infty} e^{-\frac{\pi\beta}{4}\left[(2k+1)^2 + (2n+1)^2\right]}(n+k+1)\cos\left[\vartheta^\mu_{n,k}(x,t)(k-n)\right]. \tag{A.33}$$

Thus, general solution (A.28) to inhomogeneous equation (2.4)/(2.5) has the form:

$$f^1_{\mu,\beta}\langle v \rangle^{\mu,\beta}_1 = \frac{l}{2\mu T_\mu N(\beta)} \sum_{n,k=-\infty}^{+\infty} e^{-\frac{\pi\beta}{4}\left[(2k+1)^2 + (2n+1)^2\right]}(n+k+1)T_{|k-n|}\left(\cos\left[\vartheta^\mu_{n,k}(x,t)\right]\right) + const, \tag{A.34}$$

where constant value *const* may be determined from the initial-boundary conditions. To determine it, we calculate the velocity of motion of the center of mass $\langle\langle v \rangle\rangle(t)$

$$\langle\langle v \rangle\rangle(t) = \int_0^l f^\beta_\mu(x,t)\langle v \rangle_{o.н}(x,t)dx =$$

$$\frac{l}{2\mu T_\mu N(\beta)} \sum_{n,k=-\infty}^{+\infty} e^{-\frac{\pi\beta}{4}\left[(2k+1)^2 + (2n+1)^2\right]}(n+k+1)\int_0^l \cos\left[\vartheta^\mu_{n,k}(x,t)(k-n)\right]dx + const \cdot l,$$

considering

$$\int_0^l \cos\left[\vartheta_{n,k}^\mu(x,t)(k-n)\right]dx = \frac{2}{A}\cos\left\{2\pi(k-n)\left(\mu+1-\frac{t}{T_\mu}(n+k+1)\right)\right\}\sin\left[2\pi\mu(k-n)\right] = 0,$$

we obtain

$$\langle\langle v \rangle\rangle(t) = const \cdot l. \qquad (A.35)$$

As shown in the proof of Theorem 3, dependence $n+k+1=0$ makes the main contribution $\dfrac{2}{N(\beta)}e^{-\frac{\pi\beta}{2}}\cos\left[\pi\left(2\mu\dfrac{x}{l}+1\right)\right]$ at $n=-1, k=0$ and $n=0, k=-1$ to the sum of series (A.34). Multiplying the main contribution by zero factor $n+k+1=0$ gives zero.

Theorem 7 is proved.

## Proof of Theorem 8

Using notations (3.3), let us calculate partition sum $Z(\beta)$, we obtain

$$Z(\beta) = \sum_{k=-\infty}^{+\infty} e^{-E_\mu(2k+1)^2 \frac{\pi\beta}{2E_\mu}} = \sum_{k=-\infty}^{+\infty} e^{-\pi\beta(2k+1)^2} = \theta_1\left(-\frac{1}{2}, 2i\beta\right) = \theta_1\left(-\frac{1}{2}, i\frac{4E_\mu}{\pi}\beta\right).$$

Let us consider limit (3.9). By analogue to the proof of Theorem 3, the expression for $\langle \mathcal{E}_\mu \rangle_{Gibbs}$ with an accuracy up to the terms of the first order of smallness will have the form

$$\langle \mathcal{E}_\mu \rangle_{Gibbs} \approx \frac{E_\mu}{\sum_{k=-1}^{0} e^{-\frac{\pi\beta}{2}(2k+1)^2}} \sum_{k=-1}^{0} e^{-\frac{\pi\beta}{2}(2k+1)^2}(2k+1)^2 = E_\mu.$$

Theorem 8 is proved.

## Proof of Theorem 9

Let us perform direct calculations

$$\overline{f}_{\mu,\beta}^1(x)\langle\overline{\mathcal{E}}_{\mu,\beta}\rangle(x) = \frac{E_\mu}{N(\beta)}\sum_{n,k=-\infty}^{+\infty} e^{-\frac{\pi\beta}{4}\left[(2n+1)^2+(2k+1)^2\right]}(n+k+1)^2 J_{n,k}^\mu(x), \qquad (A.36)$$

where integral $J_{n,k}^\mu(x)$ has been calculated in the proof of Theorem 4 (A.23). Substituting (A.23) into expression (A.36), we obtain

$$\overline{f}_{\mu,\beta}^1(x)\langle\overline{\mathcal{E}}_{\mu,\beta}\rangle(x) = \frac{E_\mu}{N(\beta)}\sum_{k=-\infty}^{+\infty} e^{-\frac{\pi\beta}{2}(2k+1)^2}(2k+1)^2. \qquad (A.37)$$

With the use of notations (3.1) and (3.3), expression (A.37) will take the from

$$\langle \bar{\mathcal{E}}_{\mu,\beta} \rangle(x) = \frac{2}{l} \frac{1}{Z(\beta)} \sum_{\kappa} e^{-\beta \mathcal{E}_\kappa} \sin^2(\kappa x) = \frac{1}{l} \frac{1}{Z(\beta)} \sum_{\kappa} e^{-\beta \mathcal{E}_\kappa} \mathcal{E}_\kappa = \frac{1}{l} \langle \mathcal{E}_\mu \rangle_{Gibbs}(\beta), \qquad (A.38)$$

$$\langle \bar{\mathcal{E}}_{\mu,\beta} \rangle(x) = \frac{\langle \mathcal{E}_\mu \rangle_{Gibbs}(\beta)}{2 \langle \sin^2(\kappa x) \rangle_{Gibbs}}, \quad \lim_{\beta \to +\infty} \langle \bar{\mathcal{E}}_{\mu,\beta} \rangle(x) = \frac{\lim_{\beta \to +\infty} \langle \mathcal{E}_\mu \rangle_{Gibbs}(\beta)}{2 \sin^2\left(\frac{\pi \mu}{l} x\right)} = \frac{1}{l} \frac{E_\mu}{f_\mu^1(x)}, \qquad (A.39)$$

where (3.5) is taken into consideration. Averaging expression (A.37) over the coordinate and considering (A.38), we obtain

$$\bar{f}_{\mu,\beta}^0 \langle \langle \bar{\mathcal{E}}_{\mu,\beta} \rangle \rangle = \langle \mathcal{E}_\mu \rangle_{Gibbs}(\beta),$$
$$\langle \langle \bar{\mathcal{E}}_{\mu,\beta} \rangle \rangle = \langle \mathcal{E}_\mu \rangle_{Gibbs}(\beta), \qquad (A.40)$$

where the following is taken into account

$$\bar{f}_{\mu,\beta}^0 = \int_0^l \bar{f}_{\mu,\beta}^1(x) dx = \frac{2}{l} \int_0^l \langle \sin^2(\kappa x) \rangle_{Gibbs} dx =$$
$$= \frac{2}{l} \frac{1}{Z(\beta)} \sum_{\kappa} e^{-\beta \mathcal{E}_\kappa} \int_0^l \sin^2(\kappa x) dx = 2 \frac{1}{2} \frac{1}{Z(\beta)} \sum_{\kappa} e^{-\beta \mathcal{E}_\kappa} = 1.$$

as well as expression (3.4). Theorem 9 is proved.

*Proof of Theorem 10*

Assertion (3.7) of Theorem 8 implies that

$$d\mathcal{S} = k_B d\left[ \beta \langle \mathcal{E}_\mu \rangle_{Gibbs} + \ln Z(\beta) \right], \qquad (A.41)$$

from here

$$\mathcal{S}(\beta) = k_B \left[ \beta \langle \mathcal{E}_\mu \rangle_{Gibbs}(\beta) + \ln Z(\beta) \right] + \text{const}. \qquad (A.42)$$

Let us define the following constant in expression (A.42).

$$\lim_{\beta \to +\infty} \mathcal{S}(\beta) = \lim_{\beta \to +\infty} \bar{\mathcal{S}}(\beta) = k_B \lim_{\beta \to +\infty} \left[ \frac{\pi \beta}{2E_\mu} \langle \mathcal{E}_\mu \rangle_{Gibbs}\left(\frac{\pi \beta}{2E_\mu}\right) + \ln Z\left(\frac{\pi \beta}{2E_\mu}\right) \right] + \text{const} =$$
$$= k_B \lim_{\beta \to +\infty} \left[ \frac{\pi \beta}{2} + \ln \bar{Z}(\beta) \right] + \text{const} = k_B \ln \lim_{\beta \to +\infty} \left[ \bar{Z}(\beta) e^{\frac{\pi \beta}{2}} \right] + \text{const} \qquad (A.42)$$

where (3.9) is taken into account. Let's calculate the limit

$$\lim_{\beta \to +\infty} \left[ \bar{Z}(\beta) e^{\frac{\pi \beta}{2}} \right] = \lim_{\beta \to +\infty} \sum_{k=-\infty}^{+\infty} e^{-\frac{\pi \beta}{2}\left[(2k+1)^2 - 1\right]} = \lim_{\beta \to +\infty} \sum_{k=-\infty}^{+\infty} e^{-2\pi \beta k(k+1)} = 2, \qquad (A.43)$$

where we take into account that when taking the limit in the sum (A.43), only two terms with numbers $k=0$ and $k=-1$ make a considerable contribution. Substituting (A.43) into (A.42), we obtain

$$\lim_{\beta \to +\infty} \mathcal{S}(\beta) = k_B \ln 2 + \text{const}, \qquad (A.44)$$

where, assuming $\text{const} = -k_B \ln 2$, we obtain the validity of expressions (3.19) and (3.18):

$$\mathcal{S}(\beta) = -k_B \left[ \ln e^{-\beta \langle \mathcal{E}_\mu \rangle_{Gibbs}(\beta)} + \ln \frac{2}{Z(\beta)} \right] = -k_B \ln \left[ \frac{2}{Z(\beta)} e^{-\beta \langle \mathcal{E}_\mu \rangle_{Gibbs}(\beta)} \right].$$

Let us calculate the limit (3.19) at $\beta \to 0+$.

$$\lim_{\beta \to 0+} \mathcal{S}(\beta) = -k_B \ln \left[ 2 \lim_{\beta \to 0+} \frac{e^{-\beta \langle \mathcal{E}_\mu \rangle_{Gibbs}(\beta)}}{Z(\beta)} \right] = -k_B \ln \left[ 2 \lim_{\beta \to 0+} \frac{e^{-\beta E_\mu}}{Z(\beta)} \right] = -k_B \ln \left[ 2 \lim_{\beta \to 0+} \frac{1}{Z(\beta)} \right] = +\infty,$$

where $\lim_{\beta \to 0+} Z(\beta) = +\infty$. Theorem 10 is proved.